\documentclass[fleqn,usenatbib,useAMS]{mnras}

\usepackage{graphicx}	
\usepackage{amsmath}	
\usepackage{amssymb}	
\usepackage{multicol}        
\usepackage{diagbox}
\usepackage{bm}		
\usepackage{pdflscape}	
\usepackage{journals}

\newcommand{\revone}[1]{\textcolor{black}{#1}}
\newcommand{\revtwo}[1]{\textcolor{black}{#1}}

\def\N{NGC\,}
\def\NGC{NGC\,}
\def\UGC{UGC\,}
\def\U{UGC\,}
\def\Malin{Malin\,}
\def\HI{H{\sc i}\, }

\def\iii{\,{\sc iii}}
\def\Ms{$\textrm{M}_{\odot}$}
\def\kms{$\textrm{km~s$^{-1}$}$}
\def\nb{\textsc{nbursts}}

\usepackage[T1]{fontenc}
\usepackage{ae,aecompl}
\usepackage{newtxtext,newtxmath}
\usepackage{epstopdf}

\title[On the origin of gLSBGs]{Observational insights on the origin of giant low surface brightness  galaxies}
\author[A. Saburova et al.]{
Anna S. Saburova,$^{1,2}$\thanks{E-mail:saburovaann@gmail.com}
Igor V. Chilingarian,$^{3,1}$
Anastasia V. Kasparova,$^{1}$
\newauthor
Olga K. Sil'chenko,$^{1}$
Kirill A. Grishin,$^{1,4}$
Ivan Yu. Katkov,$^{5,6,1}$
Roman I. Uklein$^{7}$
\\
$^1$ Sternberg Astronomical Institute, Moscow M.V. Lomonosov State University, Universitetskij pr., 13,  Moscow 119234, Russia\\
$^2$ Institute of Astronomy, Russian Academy of Sciences, Pyatnitskaya st., 48,  Moscow 119017, Russia\\
$^3$ Center for Astrophysics --- Harvard and Smithsonian, 60 Garden Street MS09, Cambridge, MA 02138, USA\\
$^4$ Department of Physics, M.V. Lomonosov Moscow State University, 1 Vorobyovy Gory, Moscow 119991, Russia\\
$^5$ New York University Abu Dhabi, PO Box 129188 Abu Dhabi, UAE\\
$^6$ Center for Astro, Particle, and Planetary Physics, NYU Abu Dhabi, PO Box 129188, Abu Dhabi, UAE\\
$^7$ Special Astrophysical Observatory, Russian Academy of Sciences, Nizhniy Arkhyz, Karachai-Cherkessian Republic 357147, Russia\\
}

\begin{document}
\label{firstpage}
\pagerange{\pageref{firstpage}--\pageref{lastpage}} \pubyear{2020}
\maketitle

\begin{abstract}
Giant low surface brightness galaxies (gLSBGs) with dynamically cold stellar discs reaching the radius of 130~kpc challenge currently considered galaxy formation mechanisms. We analyse new deep long-slit optical spectroscopic observations, archival optical images and published \HI and optical spectroscopic  data for a sample of seven gLSBGs, for which we performed mass modelling and estimated the parameters of dark matter haloes assuming the Burkert dark matter density profile. Our sample is not homogeneous by morphology, parameters of stellar populations and total mass, however, six of seven galaxies sit on the high-mass extension of the baryonic Tully--Fisher relation. In \U1382 we detected a global  counterrotation of the stellar high surface brightness (HSB) disc with respect to the extended LSB disc. In \UGC1922 with signatures of a possible merger, the gas counterrotation is seen in the inner disc. Six galaxies host active galactic nuclei, three of which have the estimated black hole masses substantially below those expected for their (pseudo-)bulge properties suggesting poor merger histories. Overall, the morphology, internal dynamics, and low star formation efficiency in the outer discs indicate that the three formation scenarios shape gLSBGs: (i) a two-stage formation when an HSB galaxy is formed first and then grows an LSB disc by accreting gas from an external supply; (ii) an unusual shallow and extended dark matter halo; (iii) a major merger with fine-tuned orbital parameters and morphologies of the merging galaxies.
\end{abstract}

\section{Introduction}\label{intro}

\citet{Freeman1970} discovered that the majority of galactic discs have exponential light profiles with nearly the same $B$-band central surface brightnesses of 21.65 ${\rm mag~arcsec}^{-2}$. However, over a decade later, deep photometric observations revealed the existence of fainter systems, which were named low surface brightness (LSB) galaxies. \citet{Bothun1987} discovered a separate subclass of a stellar system, giant LSB galaxies (gLSBGs) with a disc radius upto 130~kpc \citep{Boissier2016} that is nearly ten times the radius of the Milky Way. These galaxies pose a problem in the currently accepted hierarchical galaxy formation paradigm. Despite they are among the most massive known galaxies reaching dynamical masses of $10^{12}$ and $\sim10^{11}$~{\Ms} in stars, they have large-scale dynamically cold discs. It is very difficult to grow such high stellar mass in the hierarchical clustering paradigm without numerous major mergers \citep{Rodriguez-Gomez2015} that would likely overheat and destroy the discs \citep{Wilman2013}. Most gLSBGs live in a sparse environment or total isolation \citep{Saburovaetal2018}, and perhaps this can help to preserve gLSB discs through the cosmic time \citep{Galaz2011, lsb_env2019}. However, at a certain stage of the gLSBGs evolution there should exist a large reservoir of gas sufficient to foster the formation of a massive gaseous disc. 

The question `how do gLSBGs form?' remains a matter of debate \citep[see, e.g.][and references therein]{Kasparova2014,Galaz2015,Boissier2016,Hagen2016, Saburovaetal2018}. The recent studies discuss the two groups of gLSBGs formation scenarios, (i) non-catastrophic scenarios involving either slow gas accretion from filaments or secular evolution, and (ii) catastrophic models based on major- and/or minor-merger events. 

The major advantage of catastrophic scenarios from the point of view of simulations is that they could work within the current galaxy formation framework. For example, \citet{Zhuetal2018} found an analogue of \Malin1 in the IllustrisTNG simulation, which satisfactorily reproduced most observed  properties of \Malin1 and was formed by a merger of three quite massive galaxies.
\citet{Saburovaetal2018} considered a major-merger scenario among others in the case of \U1922, a gLSBG with a counterrotating inner gaseous disc. Using dedicated N-body/hydrodynamical simulations, they argued that its gLSB disc can be the result of an in-plane merger of a giant early-type spiral (Sa) galaxy and a gas-rich late-type (Sd) giant companion on a prograde orbit.
However, the counterrotation observed in the central gaseous component with respect to the outer disc of \U1922 should be the result of another (minor) merger event.
Despite relatively rich environment of \U1922 compared to other gLSBGs, \citet{Saburovaetal2018} proposed that the major-merger scenario could also be a viable option for other galaxies of this class, which have fewer known companions. 

Earlier, \citet{Mapelli2008} proposed another configuration of the catastrophic scenario that could lead to the formation of a gLSBG, a bygone head-on collision of a galaxy with a massive intruder, which could form a system similar to \Malin1 as a result of the expansion of a collisional ring. 
The weak point of this scenario is that the progenitor galaxy that experienced a collision should be already an LSB system. 
Also, \citet{Kasparova2014}, \citet{Boissier2016} and \citet{Hagen2016} did not find evidence in favour of this scenario for \Malin2, \Malin1, and \U1382.  
Instead, to explain the origin of \U1382, \citet{Hagen2016} lean towards another widely discussed formation channel proposed by \citet{Penarrubia2006}, the accretion of several gas-rich low-mass satellites. Similar scenario is also proposed as one of the channels of the formation of massive discs by \citet{Jackson2020}.

Despite the fact that most gLSBGs have diffuse clumps in their discs \citep{Kasparova2014, Boissier2016, Hagen2016,Saburovaetal2018} which can be traces of recent mergers, the minor merger scenario by \citet{Penarrubia2006} contradicts to most published H{\sc i} gLSB observations \citep[see, e.g.][]{Pickering1997, Mishra2017} because it predicts the fall-off of the rotational velocity at the periphery of the disc, which is not observed.

Along with the catastrophic formation scenarios for gLSBGs, several studies proposed non-catastrophic solutions, the scenarios which do not include a major merger or disruption of the satellites. \citet{Noguchi2001} discussed the transformation of normal HSB spirals to gLSBGs through dynamical evolution due to a bar, which induces non-circular motions and radial mixing of disc matter that flattens the disc density profile. \citet{Kasparova2014} proposed that the large radius of the disc could be related to the a `sparse' and shallow dark matter halo. They found that the dark matter halo of \Malin2 has the peculiarly high radial scale and low central density, and concluded that it could have caused the formation of a giant disc. 

\citet{Saburova2018} studied high surface brightness disc galaxies with slightly smaller (compared to gLSBGs) but still very large radii and highlighted a similar trend of larger radial scales of dark matter haloes for these systems in comparison to ``normal-sized'' disc galaxies. 
Central densities of dark matter haloes of HSB giant discs lie within the scatter for ordinary galaxies. 
Perhaps, to form a gLSBGs, both a large scale and a lower central density of the dark halo are required.

The evidence in favour of this idea were found e.g. in \citet{lsb_env2019}, who concluded that the spin parameters of LSB galaxies are systematically higher than those of HSB systems. 
If the baryons share the specific angular momentum with a dark halo \citep{Fall1980} than it could lead to larger radial scales of stellar discs and their lower baryonic surface densities in LSB galaxies, which was proved by self-consistent hydrodynamic simulations \citep[see, e.g.][]{kimlee2013}.
The properties of dark halos could, in turn, be related to the environment both at the stage of the galaxy formation and during its latter lifespan. 

Most published works on gLSBGs are devoted to individual objects primarily because such objects are very difficult to observe and study in detail. In this paper we discuss a larger sample that includes seven gLSBGs. We present new observations for the four gLSBGs: \Malin2, \N7589, \U1382, and \U6614 and compare them to the data already available in the literature for \Malin1, \U1378, and \U1922. 
We present the results of long-slit spectral observations with the Russian 6-m telescope and the 8-m Gemini-North telescope. These data fill the central gap in low-resolution profiles of internal kinematics derived from radio observations in H{\sc i} \citep[][]{Pickering1997, Mishra2017} and give important clues about the central structure of gLSBGs.  It allows us to build the most complete picture of the formation and evolution of these unusual systems up-to-date.

The paper is organized as follows. 
We describe our sample and properties of individual galaxies in Section~\ref{gen_prop}. 
The details of observations and data reduction performed in this study are given in Section~\ref{Obs}. 
We discuss the results of mass modelling of the rotation curves in Section~\ref{massmod} and give the details on it in Section \ref{appendix}. 
In Section~\ref{sr}, we discuss the star formation rates (SFRs) of gLSBGs,  their position on the baryonic Tully--Fisher (TF) relation, and in Section \ref{form_sc} we discuss the properties of  central regions of gLSBGs, propose the formation scenarios for each galaxy and compare gLSBGs with other extended LSB and giant HSB galaxies. 
Section~\ref{conclusion} summarizes our findings.

\section{Our sample of giant low surface brightness galaxies}\label{gen_prop}
\begin{figure}
 \includegraphics[width=1.0\hsize,trim={1.5cm 2.5cm 1.5cm 1.5cm}]{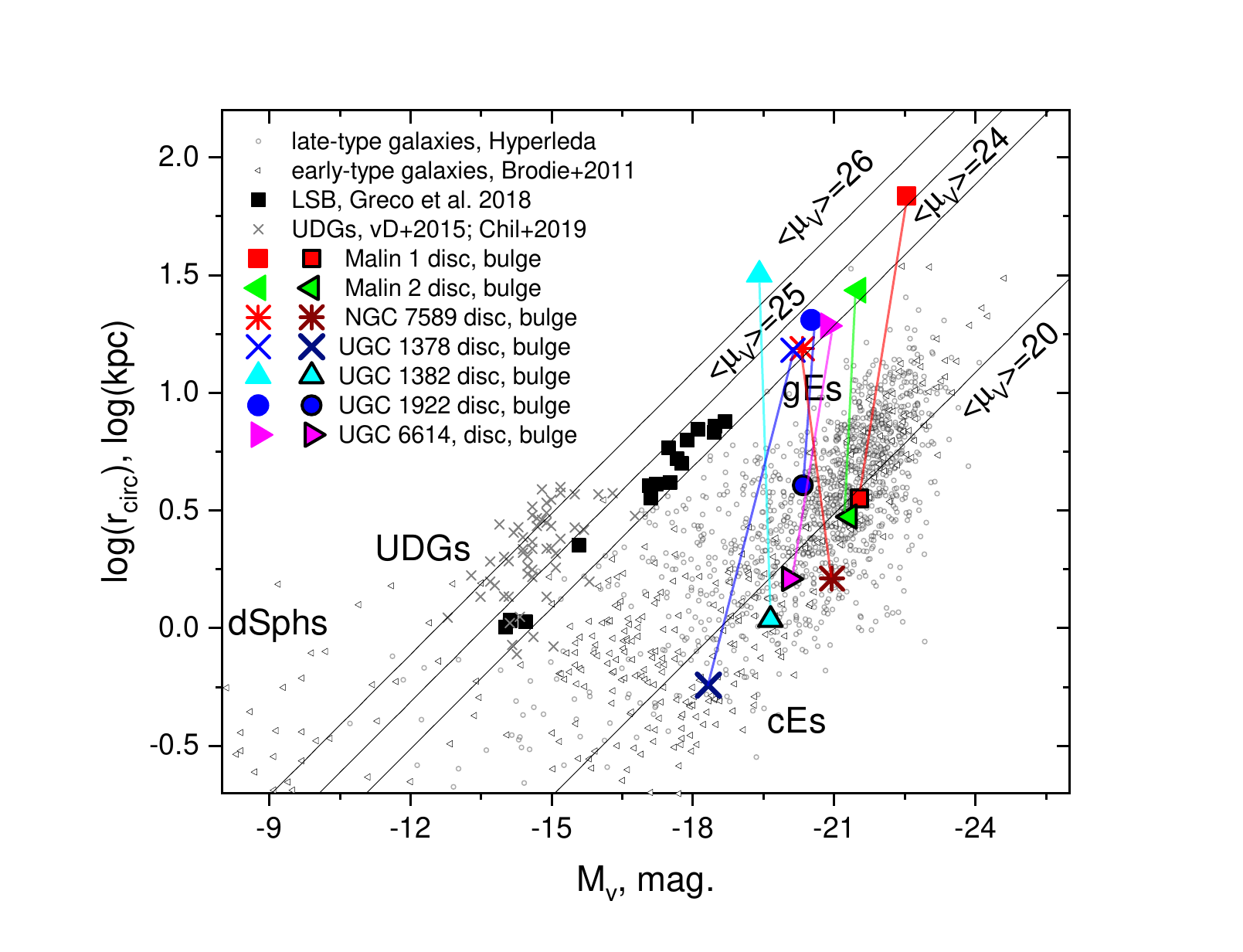}
 \caption{Position of gLSB discs and bulges on the size-luminosity relation are shown by coloured and darker outlined symbols respectively. Bulge and disc positions of the same galaxy are connected by a line. Black squares show LSB galaxies from \citet{Greco2018} with spectroscopic redshifts. Small triangles display early-type galaxies from \citet{Brodie2011}. Small circles correspond to late-type galaxies (with morphological type $t>2$) from the Hyperleda database. Small crosses show ultradiffuse galaxies from \citet{vanDokkum2015} and \citet{Chil2019}. The solid diagonal lines show three values of a constant mean surface brightness.}
\label{sizelum}
\end{figure}
\begin{figure*}
 \includegraphics[height=4.7cm, trim={0.0cm 0.2cm 0.3cm 0cm}]{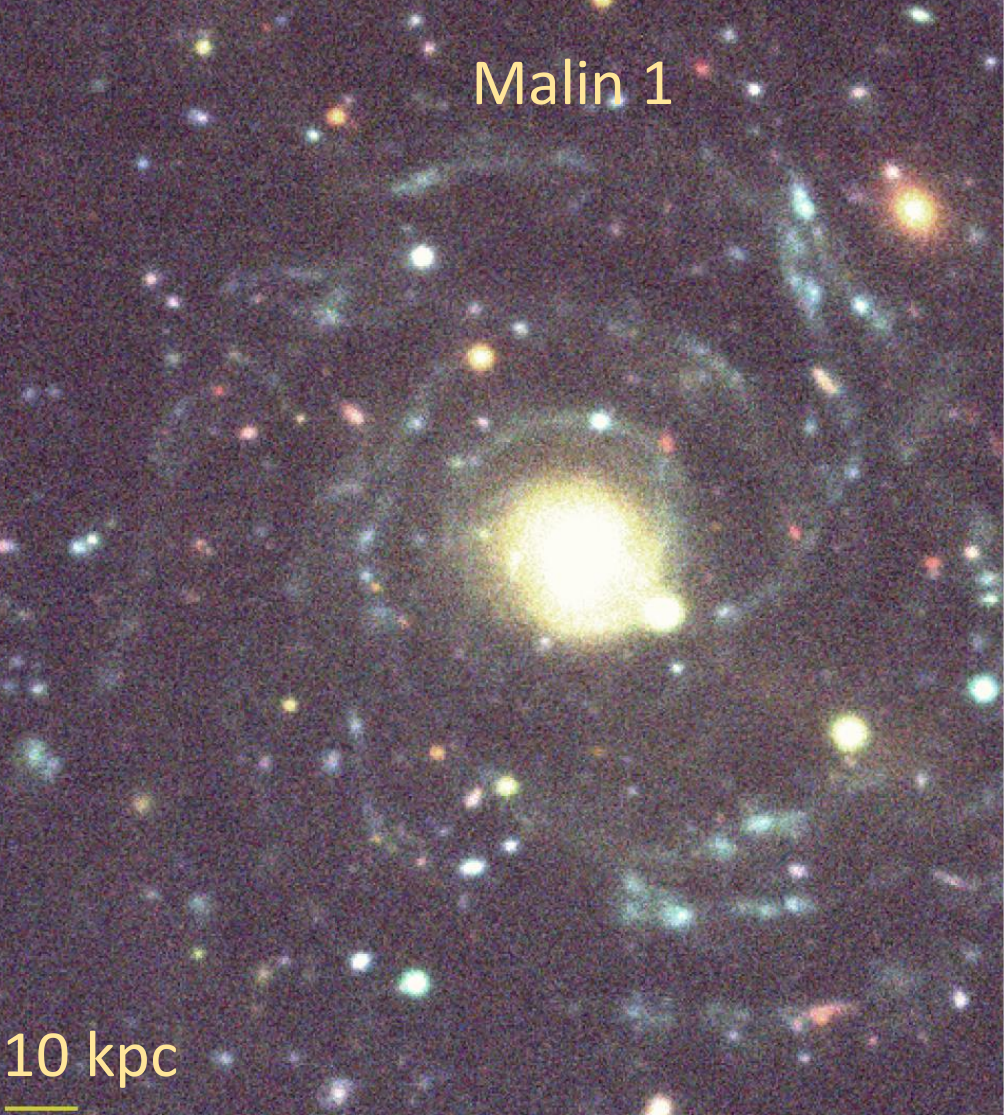}
 \includegraphics[height=4.7cm,trim={0.0cm 0.2cm 0.1cm 0cm}]{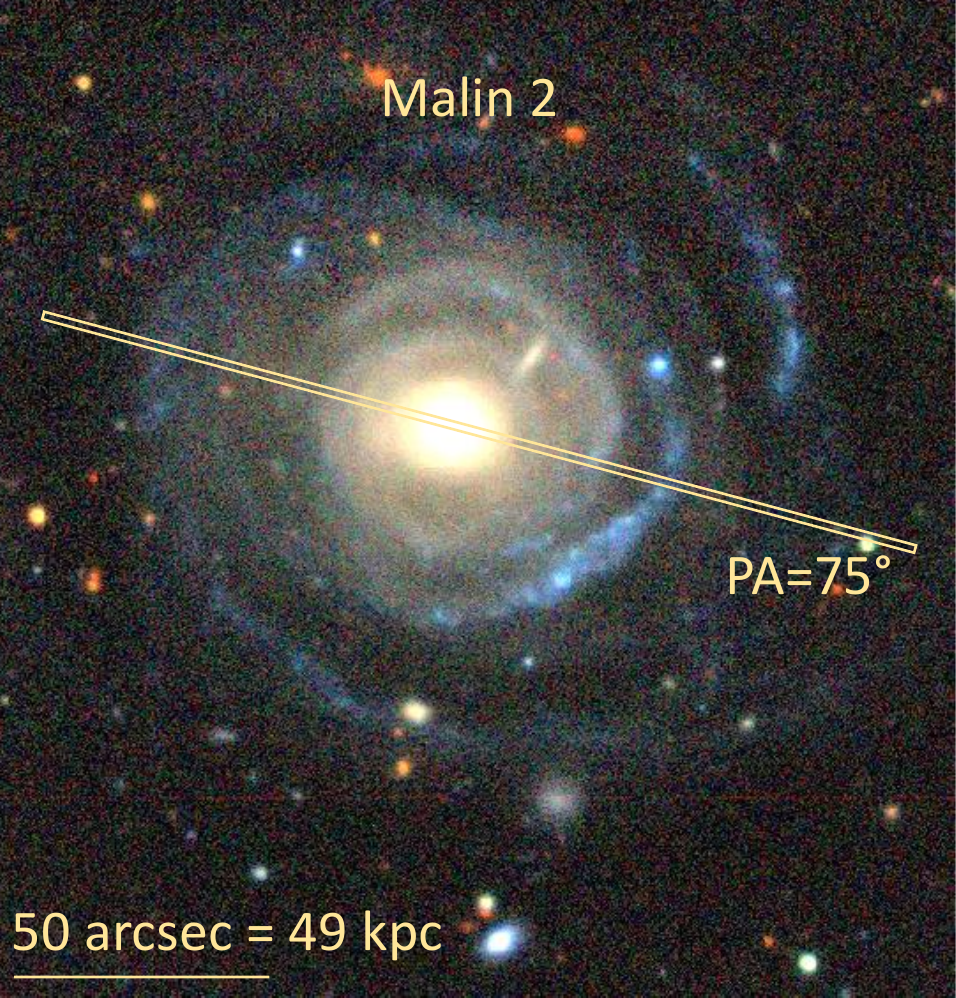}
\includegraphics[height=4.7cm,trim={0.0cm 0.2cm 0.1cm 0cm}]{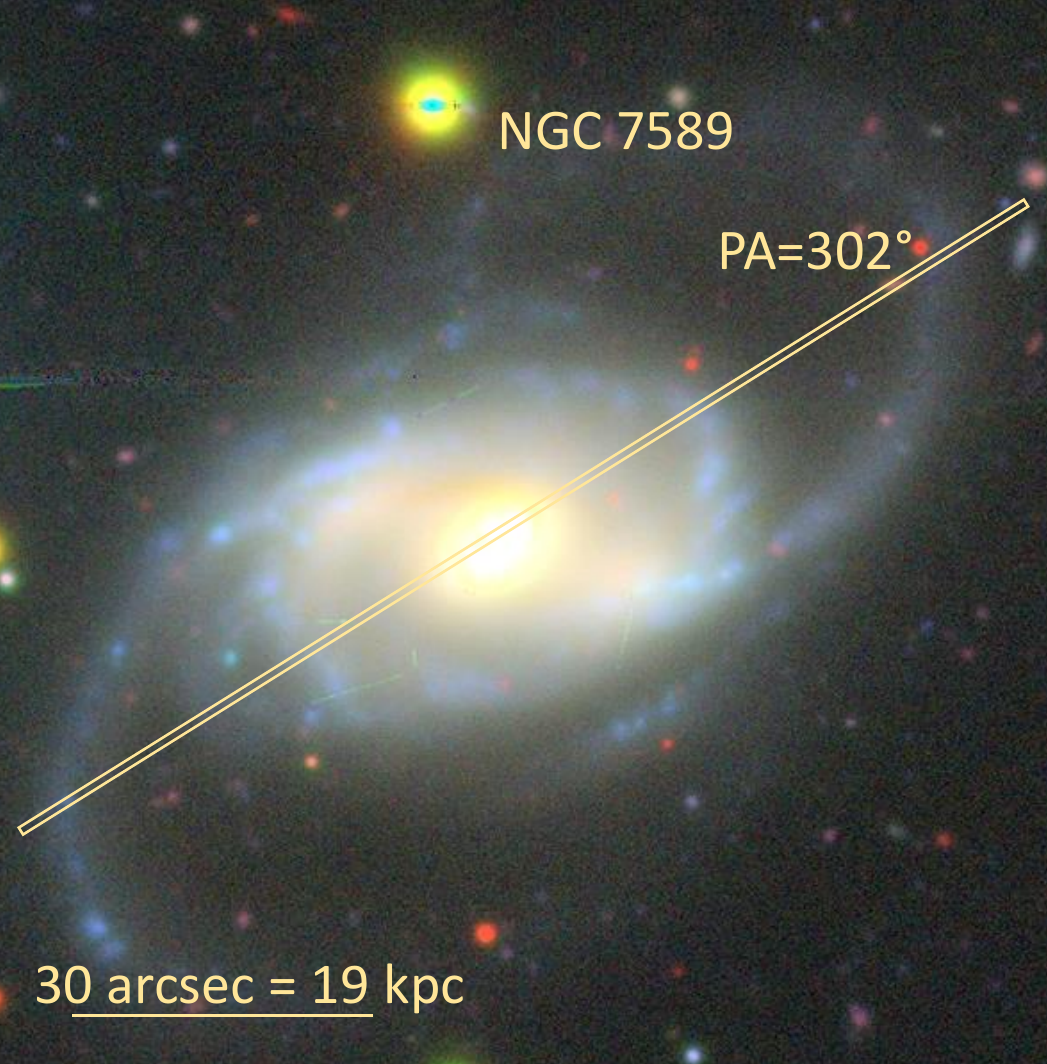}
  \includegraphics[height=4.7cm, trim={0.0cm 0.2cm 0cm 0cm}]{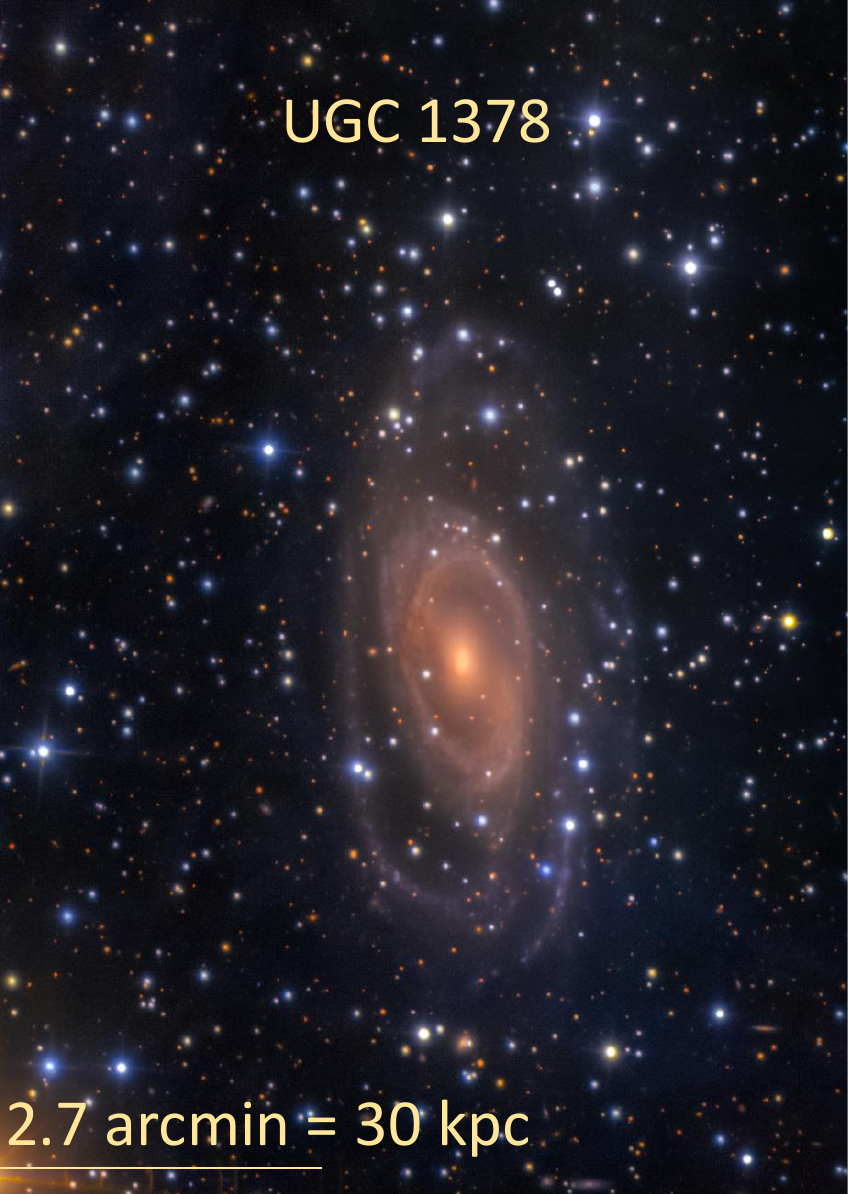}
 \includegraphics[height=4.7cm,trim={0.0cm 0.0cm 0.3cm 0cm}]{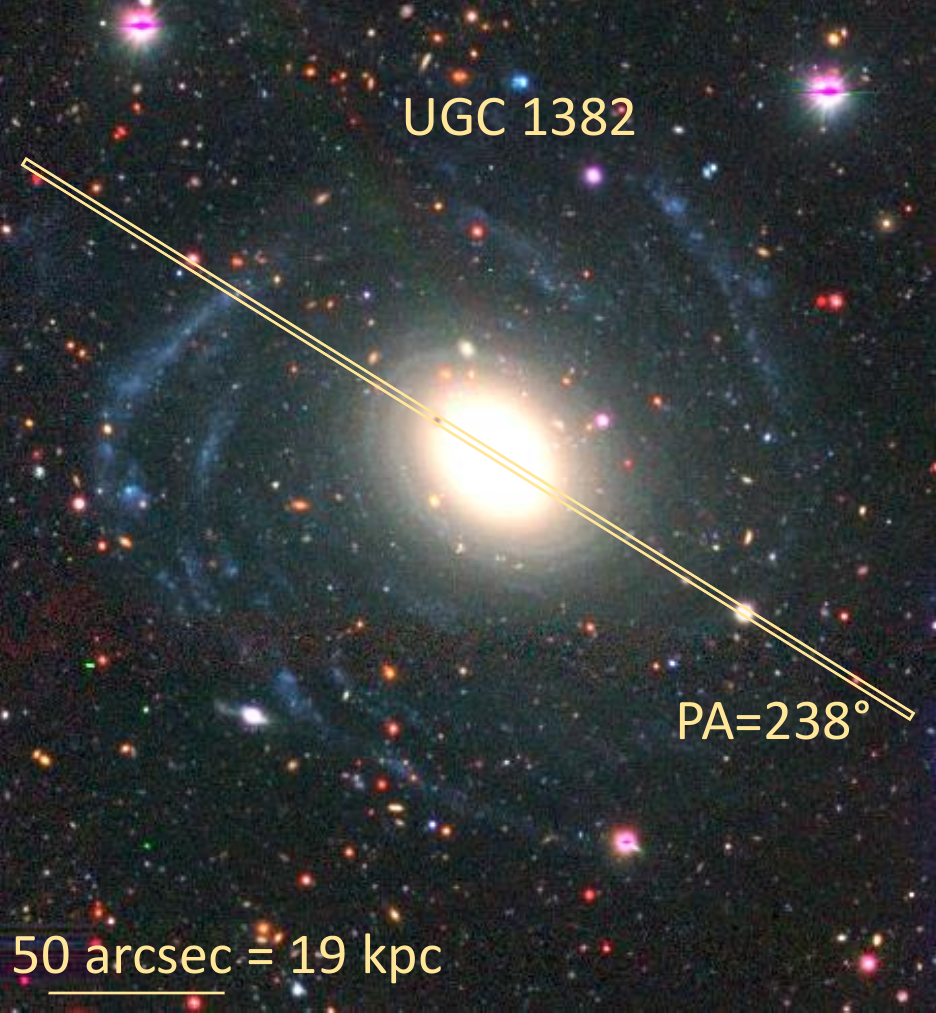}
  \includegraphics[height=4.7cm, trim={0.0cm 0.0cm 0.4cm 0cm}]{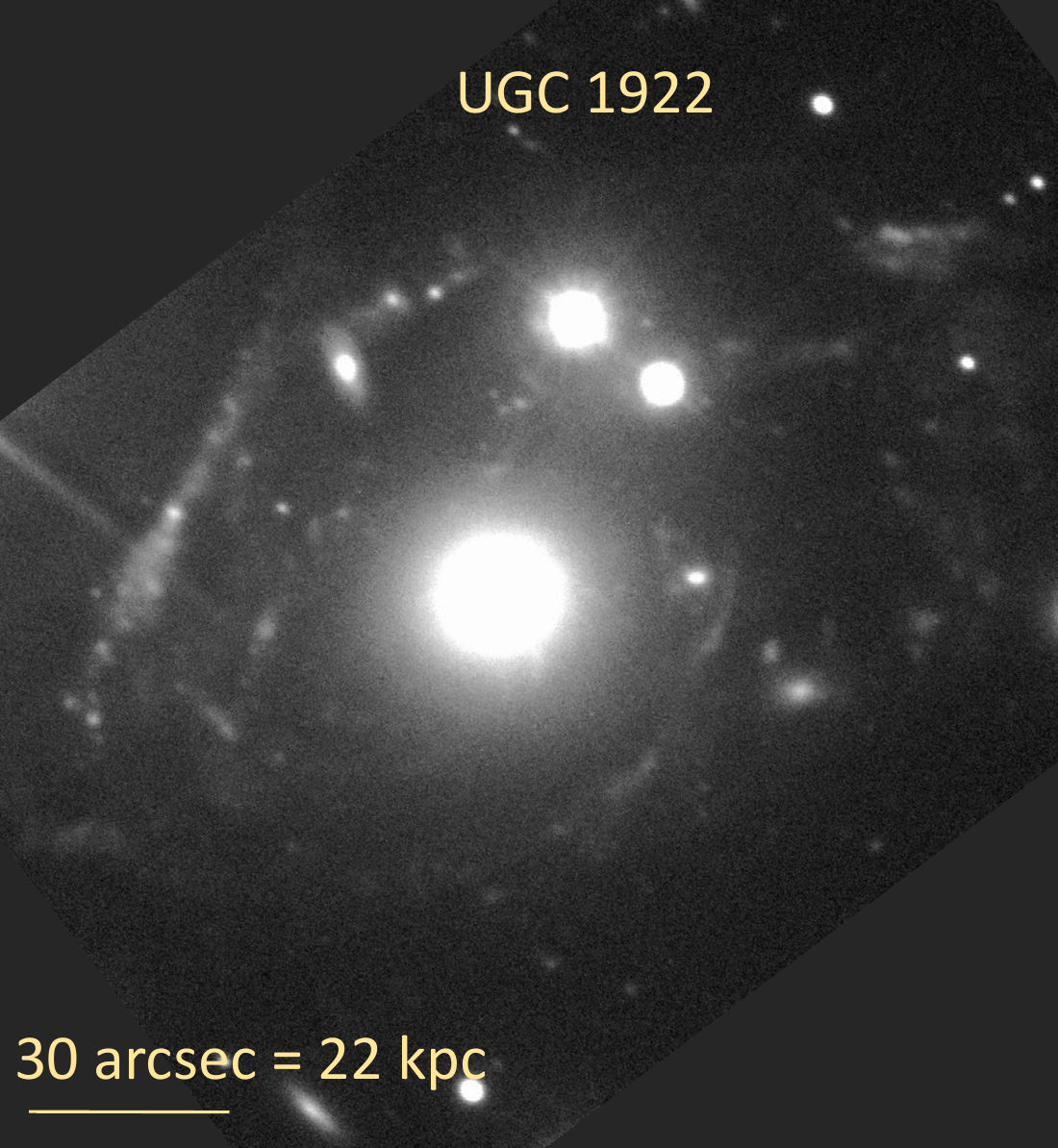}
 \includegraphics[height=4.7cm, trim={0.0cm 0.0cm 0cm 0cm}]{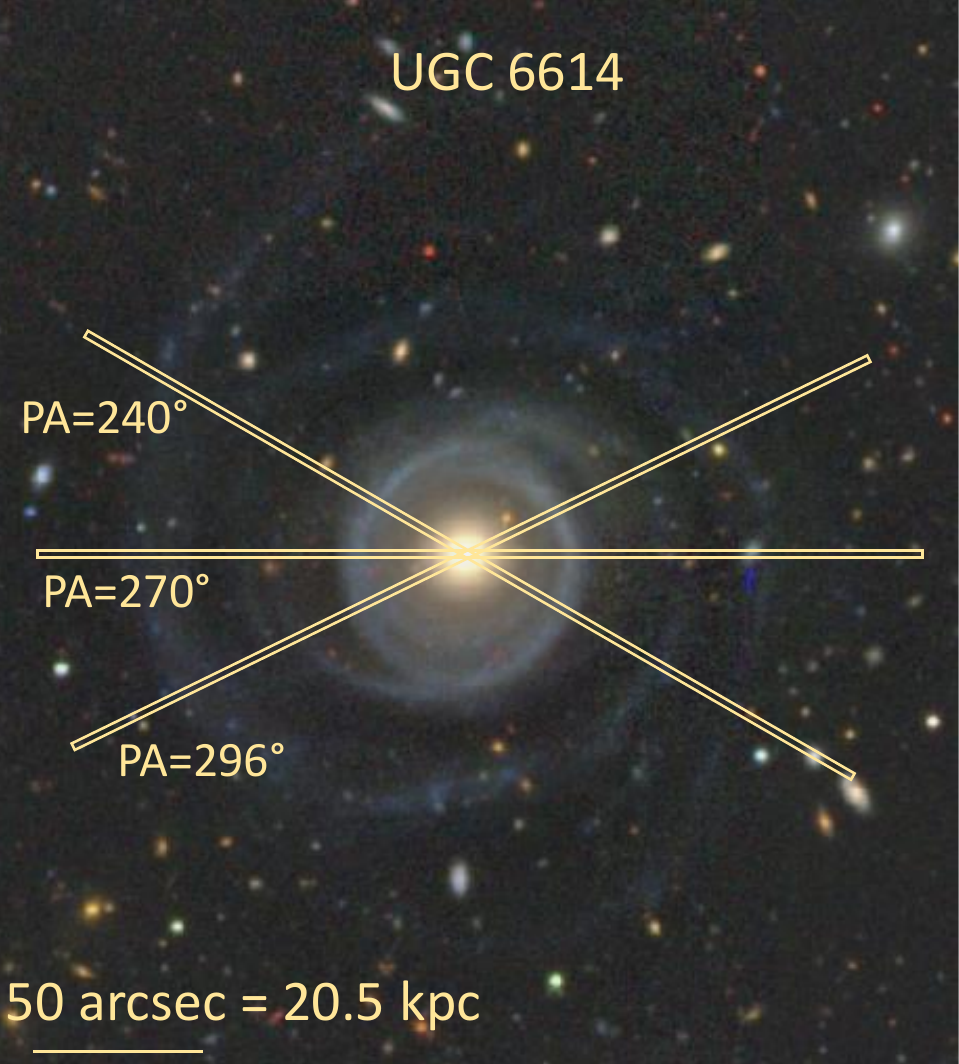}
\caption{Direct images of the seven gLSBGs from this study. The slit positions are overplotted on the images of \Malin2,  \N7589, \U1382,  and \U6614. We used {\it g, r, z} images from DECaLS \citep{Dey2019} for \Malin2 and \U6614 and from Subaru Hyper Suprime-Cam \citep{hsc2019} for \N7589 and \U1382. For Malin~1, we show a {\it u-, g-, i}-band image from the CFHT-Megacam Next Generation Virgo cluster Survey reproduced from \citet{Junais2019}. For UGC~1378 we use a {\it g-, r-, z}-band image from the MMT Binospec from \citet{Saburova2019}. For UGC~1922, we present a {\it g}-band image taken with 2.5-m telescope of the Caucasus Mountain Observatory, Sternberg Astronomical Institute from \citet{Saburova2018}.}
\label{maps}
\end{figure*}

Our sample includes seven objects. To compare them against known high- and low-surface brightness galaxies, we display them in the size--luminosity diagram (see Fig. \ref{sizelum}, coloured symbols) presenting a circularized effective radius\footnote{We calculate a circularized effective radius following \citet{Greco2018} as $r_{\mathrm{circ}}=(1-\epsilon)^{1/2}r_{\mathrm{eff}}$, where $\epsilon$ is an ellipticity and $r_{\mathrm{eff}}$ is a measured half-light radius.} versus $V$-band absolute magnitude. In some cases we converted the available $g$-band fluxes and $g-r$ colours into the $V$ band using the transformations from \citet{Jester2005}. We display bulges and gLSB discs separately. The former ones are shown with black outlines and darker colours. The black lines correspond to constant mean surface brightnesses. We also plot slightly smaller extended LSB galaxies \citep[black squares,][]{Greco2018}; ultradiffuse galaxies \citep[small crosses,][]{vanDokkum2015, Chil2019}, several families of early-type galaxies \citep[][small triangles]{Brodie2011}; galaxies with morphological types later than 2 (i.e. discs) from the Hyperleda data base (small circles). From the sample of \citet{Greco2018}, we took only the objects with available spectroscopic redshifts.

From Fig. \ref{sizelum}, it is evident that LSB discs of gLSBGs have similar mean surface brightnesses to  LSBs extending their locus to higher luminosities and also for a given luminosity they are much more extended compared to `normal' late-type galaxies from Hyperleda. At the same time, bulges of gLSBGs are similar to those in HSB early- and late-type galaxies.

We present the general properties of the gLSBGs from our sample in Tables~\ref{sample} and \ref{tabproperties}.  Below we briefly describe every individual system.

{\bf \Malin1} is the prototype of the gLSBG class discovered by \citet{Bothun1987}. Similar to most gLSBGs, it has a prominent bulge and faint extended spiral arms. It shows signs of an active galactic nucleus (AGN) lying in the borderline of the LINER--Seyfert classification by emission-line ratios \citep{Junais2020}.
Deep images of the galaxy revealed the well pronounced low surface brightness spiral arms \citep{Galaz2015, Boissier2016} while the central part resembles a ``normal'' barred early-type spiral galaxy (SB0/a) with a bulge and an HSB disc \citep{Barth2007, Saha2021}.   
\citet{Reshetnikov2010} presented spectroscopic evidence that \Malin1 is interacting with the small companion \Malin1B. 
They concluded that the interaction could lead to the appearance of a single dusty LSB spiral arm in \Malin1. 
The large-scale environment of \Malin1 is of low density. 
Based on multiband images of \Malin1, \citet{Boissier2016} conclude that its extended disc has been forming stars with a low star formation efficiency for several Gyrs.

{\bf \Malin2} is relatively well studied since its discovery by \citet{Bothun1990}. 
It has a prominent bulge and a gLSB disc with a spiral structure. 
According to \citet{Ramya2011} \Malin2 shows AGN activity. 
The estimated mass of the central black hole, $2.9\times 10^5$ \Ms~\citep{Ramya2011} appears to be lower than expected for the observed stellar velocity dispersion. However, the H$\alpha$ line profile decomposition done using the technique presented in \citet{2018ApJ...863....1C} does not reveal a broad-line component hence questioning the black hole mass estimate from \citet{Ramya2011}.
\citet{Das2010} revealed the presence of extended molecular gas in the disc of \Malin2 with the mass in the range from $4.9\times10^8$ ~to $8.3\times 10^8$~\Ms. 
The observed ratio of molecular to atomic hydrogen surface density is significantly higher than that expected in normal galaxies for the observed low value of the turbulent gas pressure and the total gas density.  
According to \citet{Kasparova2014}, one possible explanation for this imbalance is the high content of undetected cold gas (dark gas) in \Malin2. 
The fact that \Malin2 is bright in the NUV indicates the ongoing star formation in the system. 
The SFR estimate based on the NUV flux is 4.3~\Ms~yr$^{-1}$ \citep{Boissier2008}. \citet{Kasparova2014} also give the measurement of the SFR surface density in the disc of \Malin2 $\sim 2.5\times 10^{-4}$~\Ms~yr$^{-1}$kpc$^{-2}$ based on the archival GALEX data.

{\bf \N7589} was mentioned for the first time by \citet{Sprayberry1995}. 
This galaxy has spiral arms, two bars and a ring. 
According to \citet{Lelli2010}, \N7589 consists of an HSB central part and a gLSB disc.
The available \HI rotation curve exhibits a plateau at about 200~{\kms} from the radius of 5.9~kpc, there is no \HI data in the  inner part \citep{Lelli2010}.

\N7589 is a Seyfert~1 galaxy with the mass of the central black hole estimated at $9.44\times10^6$~\Ms\ \citep{Subramanian2016}. 
The upper limit of the molecular hydrogen mass is $8.25\times10^8$~\Ms\, which corresponds to the low ratio of the molecular-to-atomic hydrogen of 0.081 \citep{Cao2017}.
However, the global SFR is not very low: 1.00~\Ms~yr$^{-1}$ \citep[deduced from the NUV flux,][]{Cao2017} or 0.73 and 1.14~\Ms~yr$^{-1}$  \citep[determined from far-ultraviolet (FUV) and near-ultraviolet (NUV) luminosities by][]{Boissier2008}.

{\bf \U1378} was classified as a gLSB galaxy by \citet{Schombert1998}. 
\citet{Saburova2019} studied it in details using long-slit spectral observations and deep multiband optical photometry. 
Similarly to \Malin1 it has a complex morphology that includes an Milky way-sized central part with a bulge, bar and an HSB disc that is immersed in a large low surface brightness disc. The global SFR based on the infrared data is between 1.2 and 2.3~\Ms~yr$^{-1}$ \citep{Saburova2019}.
At the same time, the SFR surface density of the LSB disc appears to be lower than expected for the given gas surface density, which can indicate the presence of accretion. 
No CO(1-0) emission was detected from the disc of this galaxy.

{\bf \U1382} had been misclassified as an elliptical galaxy before \citet{Hagen2016} noticed that it had an extended spiral structure visible in UV images obtained by Galaxy Evolution Explorer \citep[\rm{GALEX},][]{Galex}. 
Further analysis of the multiwavelength data revealed that it appears to be a gLSB with HSB bulge+disc surrounded by a gLSB disc. 
The global SFR of \UGC1382 is 0.42~\Ms~yr$^{-1}$ and the SFR surface density of \U1382 is similar to that of the outer regions of spiral galaxies which is typical for low efficiency of star formation \citep{Hagen2016}. 
\U1382 lives in the low-density environment, at the same time \citet{Hagen2016} also discovered a possible remnant of a satellite embedded in its LSB disc.

{\bf \U1922} was erroneously classified as an elliptical galaxy too \citep[see, e.g.][]{Huchra2012}.  
\citet{Schombert1998} revealed the extended gLSB disc in it. 
\citet{Saburovaetal2018} performed deep long-slit and photometric observations of this galaxy and discovered the presence of a kinematically decoupled central component, which counter-rotates with respect to the main disc of the galaxy. 
The deep photometry  of \U1922 revealed the asymmetric spiral structure with ``rows'' and irregular star formation on the NW-side. 
The disc appeared to be strongly dynamical overheated. 
Unlike many other gLSB galaxies \U1922 is not isolated but a member of a group that includes seven members \citep{Saulder2016}. 

{\bf \U6614} was initially studied by \citet{Schommer1983}. 
It has a prominent bulge, spiral arms, and a ring. 
Like \N7589, it has an AGN and can be classified as LINER \citep{Subramanian2016}. 
The nucleus of \U6614 is bright in X-ray, optical and radio frequencies. 
The large amplitude, short time-scale of X-ray variability of \U6614 can be indication of active intense accretion onto the central black hole \citep{Naik2010}. 
The estimate of the central black hole mass is $4.44\times10^6$~\Ms, which appears to be lower than expected for observed stellar velocity dispersion similarly to that of \Malin2, that can indicate that it is not in co-evolution with the host galaxy bulge \citep{Subramanian2016}. 
The  CO(1-0) emission  was detected from the disc of \U6614 and the molecular gas traces its spiral arms \citep{Das2006}. 
The corresponding estimate of the mass of molecular gas in the disc is $2.8 \times 10^8$~\Ms. 
The SFR determined from the infrared emission is 0.88~\Ms~yr$^{-1}$ \citep{Rahman2007}. We obtained the higher value of the global SFR based on GALEX FUV data and SFR vs UV--luminosity relation from \citet{Kennicutt1998}: 2.24~\Ms~yr$^{-1}$ and 1.44 for peripheral LSB regions. Our global SFR estimate is also higher than that derived by \citet{Wyder2009} from NUV data (1.95~\Ms~yr$^{-1}$). The colour $FUV-NUV=1.01$~mag of \UGC6614 is higher than what is usually observed in LSB galaxies according to \citet{Wyder2009}. Together with a relatively low value of the SFR surface density of $2.14\times 10^{-4}$~\Ms~yr$^{-1}$kpc$^{-2}$ \citep{Wyder2009},  $3.17\times 10 ^{-4}$~\Ms~yr$^{-1}$kpc$^{-2}$ (current work), the red $FUV-NUV$ colour can indicate the absence of large amounts of stars younger than 10$^{8}$ yr. 
\U6614 contains a noticeable amount of dust: $2.6\times10^8$~\Ms, the dust-to-gas mass ratio is 0.01 \citep{Rahman2007} which is more than ten times higher than the typical value in spiral galaxies \citep{Bettoni2003}.
Another interesting detail observed in \U6614 is the blueshifted ionized gas emission in H$\alpha$ that could indicate a jet or an accretion disc hot spot along the line of sight \citep{Ramya2011}.

\begin{table} 
\caption{Equatorial coordinates (J2000.0) from NASA/IPAC Extragalactic Database (NED)$^a$ for the gLSBGs from our sample.  \label{sample}}
\begin{center}
\begin{tabular}{lll}
Galaxy&RA& Dec\\
\hline
\Malin1& 12$^h$36$^m$59$^s$.350&+14$^d$19$^m$49$^s$.32\\
\Malin2& 10$^h$39$^m$52$^s$.483&+20$^d$50$^m$49$^s$.36\\
\N7589&23$^h$18$^m$15$^s$.668&+00$^d$15$^m$40$^s$.19\\
\U1378&01$^h$56$^m$19$^s$.24&+73$^d$16$^m$58$^s$.0\\
\U1382&01$^h$54$^m$41$^s$.042& --00$^d$08$^m$36$^s$.03\\
\U1922& 02$^h$27$^m$45$^s$.930&+28$^d$12$^m$31$^s$.83\\
\U6614&11$^h$39$^m$14$^s$.872&+17$^d$08$^m$37$^s$.21\\
\hline 
\multicolumn{3}{l}{$^a$ \url{https://ned.ipac.caltech.edu/}}\\
\end{tabular}
\end{center}

\end{table}

\begin{table*} 
\caption{Basic properties of the sample of gLSBGs: name;  radius of LSB disc (four disc radial scale lengths for all galaxies except \Malin1 for which we used the distance to the furthermost measured points from the centre above the noise level and with an approximate exponential radial distribution of surface brightness; morphological type; mass of \HI; adopted distance; inclination angle to the line of sight; position angle; rotation velocity; radius of the $B$-band 25-mag isophote from Hyperleda$^b$; AGN activity flag: L=LINER, S=Seyfert; central black hole mass; global SFR. \label{tabproperties}}
\begin{center}
\renewcommand{\arraystretch}{1.3}
\begin{tabular}{llcccccccccc}
\hline
\hline 
Galaxy& $R_{\rm{LSB}}$& T &$M_{\rm H{\sc I}}$ & D& $i$& PA& v&$R_{25}$&AGN&$M_{BH}$&SFR\\
&kpc&&$10^{10}$ \Ms &Mpc&($^\circ$)&($^\circ$)&(\kms)&(kpc)&flag&$10^{6}$ \Ms&(\Ms yr$^{-1}$)\\
\hline
\Malin1 & 130$^2$&SBab$^3$ &6.7$\pm 1.0^4$&377$\pm 8^4$&38$\pm 3^{4}$&10-60$^{4}$&236$\pm 9.4^5$&11&L$^{13}$&3.63$^{+0.84}_{-0.81}$ $^{13}$&1.2--2.5$^{14}$\\
\Malin2 & 82$^6$&Scd$^3$&3.6$\pm$0.4$^7$&201$^8$&38$\pm3$$^7$&75$\pm3^7$&320$\pm7$$^7$&33&S2$^{15}$&0.29$^{+0.32}_{-0.20}$ $^{16}$&4.3$^{14}$\\
\N7589 &56$^4$&SABa$^3$&1.5$\pm0.3^4$&130$\pm8^4$&58$\pm3^7$&302$\pm4^7$&205$\pm3^4$&18&S$^{13}$&9.44$^{+1.35}_{-1.24}$ $^{13}$&0.7--1.1$^{14}$\\
\U1378&50$^9$&SBa$^3$&1.2$\pm 0.2^{10}$&38.8$^{10}$&59$\pm 5^{10}$&181$\pm 6^{10}$&282$\pm 11^{10}$&15&--&--&1.2--2.3$^9$\\
\U1382 &80$^{11}$&S0$^{11}$&1.7$\pm 0.1^{11}$&80$^{11}$&46$^{11}$&58$^c$&280$\pm 20 ^{11}$&13&L$^{19}$&--&0.42$^{+0.30}_{-0.17}$ $^{11}$\\
\U1922  &84$^{12}$&S?$^1$&3.2$\pm0.4^{10}$&150$^{10}$&51$\pm2^{10}$&128$\pm3^{10}$&432$\pm12^{10}$&18&L$^{16}$&0.39$^{+0.18}_{-0.15}$ $^{16}$&2.2$^{18}$\\
\U6614  &54$^7$&Sa$^3$&2.5$\pm0.2^7$&85$^7$&35$\pm3^7$&296$\pm3^7$&228$\pm12^7$&16&L$^{13}$&4.44$^{+0.63}_{-0.58}$ $^{13}$&0.9$^{17}$--2.24$^{18}$\\
\hline
&&&&&&&&\\
\multicolumn{12}{l}{Reference -- $^1$ NED, $^2$ \citet{Boissier2016},$^3$ Hyperleda, $^4$ \citet{Lelli2010}, $^5$ \citet{Moore2006}, $^6$ \citet{Kasparova2014}, $^7$ \citet{Pickering1997},}\\
\multicolumn{12}{l}{$^8$ \citet{Das2010}, $^9$ \citet{Saburova2019}, $^{10}$ \citet{Mishra2017}, $^{11}$ \citet{Hagen2016}, $^{12}$ \citet{Saburova2018}, $^{13}$ \citet{Subramanian2016}, }\\
\multicolumn{12}{l}{$^{14}$ \citet{Boissier2008}, $^{15}$ \citet{Schombert1998}, $^{16}$ \citet{Ramya2011}, $^{17}$ \citet{Rahman2007}, $^{18}$ current paper, $^{19}$ \citet{Chilingarian2017}.}\\
\multicolumn{12}{l}{$^b$ HyperLeda database \citep{Makarov2014}: \url{http://leda.univ-lyon1.fr/}}\\
\multicolumn{12}{l}{$^c$ according to Fig. 3 in \citep{Hagen2016}}\\
\end{tabular}
\end{center}

\end{table*}

\section{Spectroscopic and photometric data}\label{Obs}
\subsection{Spectroscopic observations and data reduction}
\subsubsection{Observations with the Russian 6-m telescope}
Here, we present long-slit spectroscopic observations of four gLSBGs conducted with the SCORPIO universal spectrograph \citep{AfanasievMoiseev2005} at the prime focus of the Russian 6-m BTA telescope. 
We present the observing log in Table~\ref{log}, which contains the position angle of the slit, date of observation, total integration time and the atmospheric seeing quality.
We utilized the VPHG2300G volume phase holographic grism that yields the spectral resolving power of $R=2400$ (full width at half-maximum~$\sim2.2$~\AA) in the wavelength range $4800<\lambda<5570$~\AA\ sampled by 0.38~\AA~pixel$^{-1}$. 
The plate scale along the 6-arcmin long, 1-arcsec-wide slit is 0.36~arcsec pixel$^{-1}$. 
In Fig.~\ref{maps}, we show direct images of gLSBGs included in our sample. For the four galaxies with new long-slit observations presented here (\Malin2, \U6614, \NGC7589, and \U1382), we also overplot the slit positions.

The spectrosopic data reduction for SCORPIO using our \textsc{idl}-based pipeline is described in details in \citet{Saburovaetal2018}. 
It includes a bias subtraction and overscan clipping, flat-field correction, the wavelength calibration using arc lines\footnote{To improve the wavelength solution accuracy we took arc spectra every 2~h and used them to reduce the corresponding science frames.}, cosmic-ray hit removal, linearization, co-adding; the night sky subtraction using the algorithm described in \citet{KatkovChilingarian2011}  and flux calibration using the spectrophotometric standard stars \emph{Feige~34}, \emph{BD~33+2642},  \emph{Feige~110}.

We took into account the instrumental line-spread function of the spectrograph along the slit and across the wavelength range, which we determined by fitting the twilight sky spectrum observed during the same night with an $R=10000$ Solar spectrum using the {\sc ppxf} full spectum fitting technique \citep{CE04}. 
We then fitted the galaxy spectra using intermediate-resolution ($R=10000$) PEGASE.HR~\citep{LeBorgneetal2004} simple stellar population  (SSP) models computed for the Salpeter initial mass function (IMF) \citep{Salpeter1955} convolved with the instrumental line-spread function of SCORPIO using the \nb{} full spectral fitting technique  \citep{Chilingarian2007a, Chilingarian2007b}.  As the result of the procedure, we obtained the best-fitting parameters of an SSP model, that is age T~(Gyr) and metallicity [Fe/H]~(dex) of stellar population. The line-of-sight velocity distribution (LOSVD) of stars was parametrized by the Gauss--Hermite function until the fourth order \citep[see][]{vanderMarel1993}. 
The resulting LOSVDs are characterized by the line-of-sight velocity, velocity dispersion and Gauss--Hermite moments $h_3$ and $h_4$, which reflect the deviation of an LOSVD from the pure Gaussian profile. 

We also analyzed the emission spectra which we obtained by subtracting the best-fitting stellar population templates from observed spectra. 
We fitted emission lines by a single Gaussian profile and derived the velocity and velocity dispersion of ionized gas also taking into account the instrumental line spread function.
\begin{table}
\caption{Observing log for SCORPIO.}\label{log}
\begin{center}
\begin{tabular}{lcccc}
\hline\hline
Galaxy &Slit PA & Date & Exposure time& Seeing \\
 &  ($^\circ$)  &  &     (s) &        (arcsec) \\
\hline
\Malin2&75&13.03.2018&10800&1.3\\
\N7589&302&21.09.2017&7200&1.4 \\
\U1382&238&07.10.2016&7200&1.5\\
\U6614&296&15.03.2018&9000&2.0 \\

\hline
\end{tabular}
\end{center}
\end{table}
\subsubsection{Gemini-North observations}

For \UGC6614 we also found deep spectroscopic data in the Gemini science archive\footnote{\url{http://archive.gemini.edu/}}. The galaxy was observed in the two different programs with the GMOS-N spectrograph operated at the 8-m Gemini-North telescope (programs GN-2006B-Q-41, P.I.: C. Onken and GN-2005B-Q-61, P.I.: L. Ferrarese). The details of GMOS-N observations are given in Table \ref{loggmos} where we list the position angles of the slit, the dates of observation, total integration time, slit widths, gratings, wavelength range, and spectral resolution for each data set. 

We used our own {\sc idl}-based data reduction pipeline for GMOS data presented in \citep{Francis+12}.
The data reduction for the GMOS-N observations obtained in the framework of the program GN-2006B-Q-41 was identical to those for Malin~2 presented in \citet{Kasparova2014}. Data reduction for intermediate-resolution B1200 data from the program GN-2005B-Q-61 was done in a similar fashion. We updated our data reduction pipeline to handle low-resolution R400 grating data, which turned out to be very useful for tracing faint H$\alpha$ emission in the LSB disc of \UGC6614. Unfortunately, because of the chip gaps in GMOS-N and slightly different wavelength ranges used in the two programs, the B1200 dataset from GN-2006B-Q-41 had H$\beta$ missing, and the B1200 data set from GN-2005B-Q-61 had [O{\sc iii}] missing.

We analysed the GMOS-N data using \nb{} in a similar way to the SCORPIO observations described above.

\begin{table*}
\caption{Observing log for \UGC6614 (GMOS-N)}\label{loggmos}
\begin{center}
\begin{tabular}{lccccccc}
\hline\hline
Program ID &Slit PA & Date & Exposure time& Slit width&Grating&Wavelength range &Spectral resolution\\
 &  ($^\circ$)  &  &     (s) &        (arcsec)&&\AA& \\
\hline
GN-2006B-Q-41&240&24.01.2007&5400&0.5& B1200+G5301&4672--6071&3400\\
GN-2005B-Q-61&270&26.12.2005&13200&0.75&B1200+G5301&4476--5800&2800\\
GN-2005B-Q-61&270&29.11.2005&4800&0.75&R400+G5305&3900--7864&800\\
\hline
\end{tabular}
\end{center}
\end{table*}

\subsection{Results of the analysis of spectroscopic data.}\label{Res}
\begin{figure*}
\includegraphics[width=0.8\linewidth]{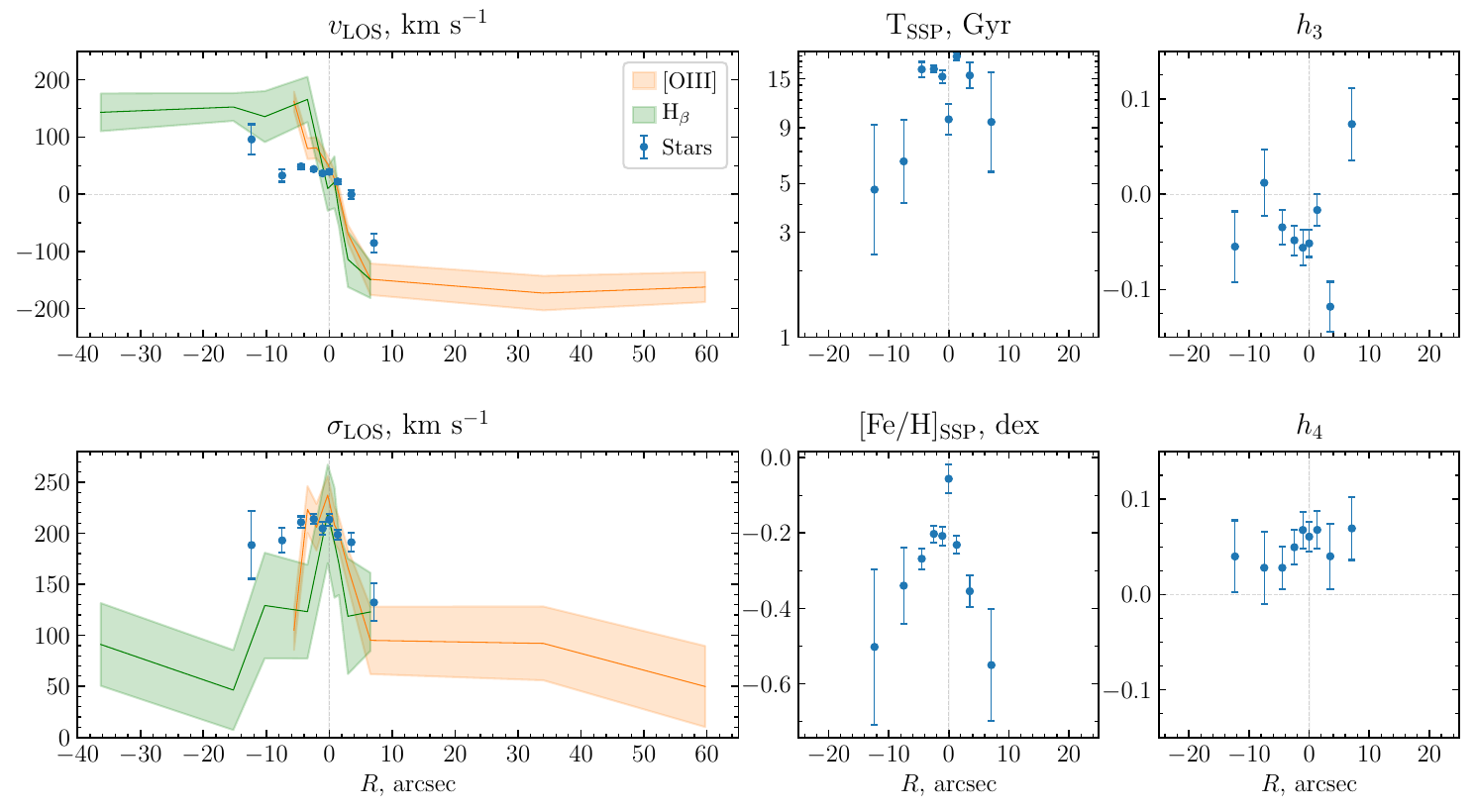}

\caption{Analysis of spectral observations of \Malin2. Left-hand panel: radial profile of the line-of-sight velocity $v$ (upper panel) and velocity dispersion $\sigma$ (bottom panel) of ionized gas (lines with shaded uncertainty areas) and stars (circles). Centre panel: radial profiles of stellar age (top panel) and metallicity (bottom panel). Right-hand panel: radial profiles of $h3$ (top panel) and $h4$ (bottom panel).}
\label{profiles_malin2}
\end{figure*}

In Fig.~\ref{profiles_malin2}, we demonstrate the main results of the analysis of spectral data for \Malin2. The left-hand column corresponds to the profiles of velocity and velocity dispersion for ionized gas (shaded lines) and stars (circles). 
The central column shows age and metallicity of stellar population. 
The right-hand column gives the profiles of $h_3$ and $h_4$ for the stellar LOSVD. 
The stars do not show clear rotation in the inner region. 
 
The trend of stellar metallicity is very close to that found by \citet{Kasparova2014}~-- the decreasing radial gradient from almost solar metallicity in the centre. 
The age of stellar population is very old in the bulge region. 
The values of $h_3$ and $h_4$ are close to zero. 

\begin{figure*}
\includegraphics[width=0.9\linewidth]{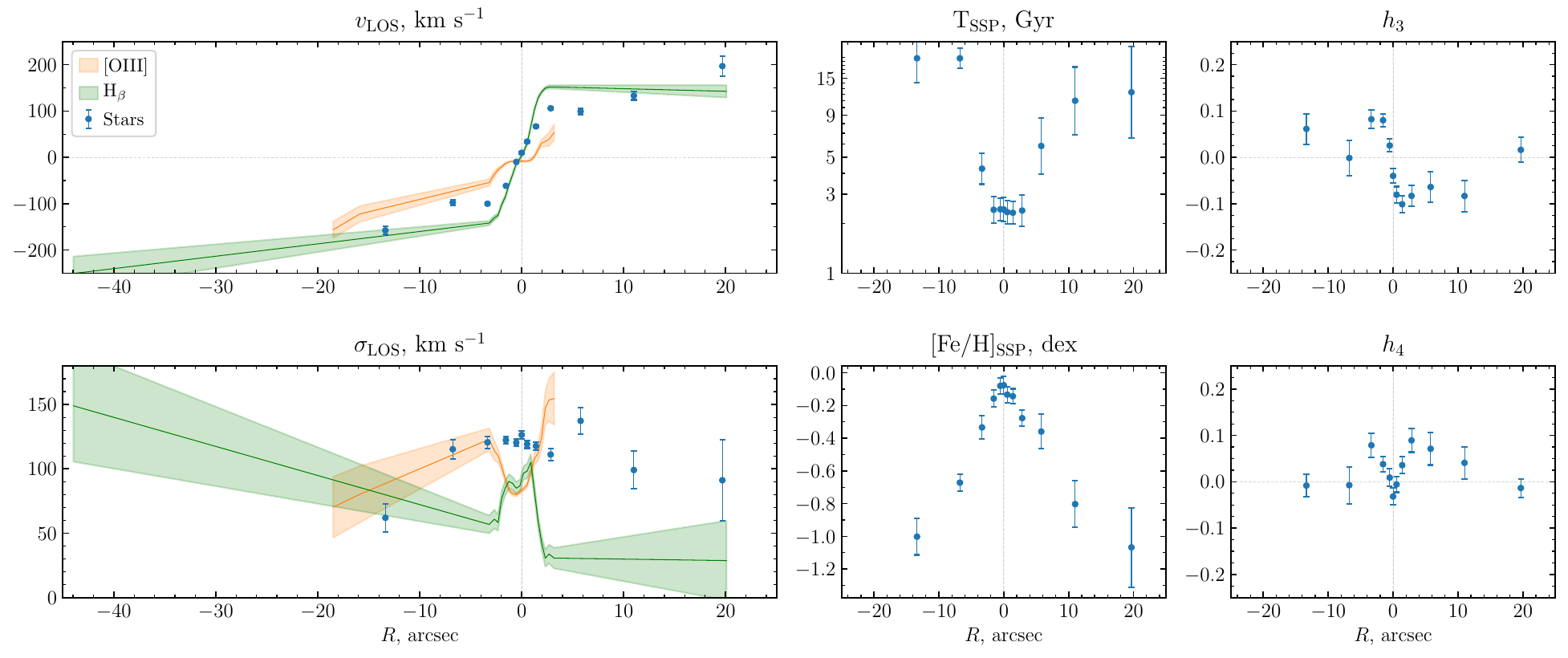}

\caption{Analysis of spectral observations of \N7589. The designations are the same as in Fig.~\ref{profiles_malin2}.}
\label{profiles_n7589}
\end{figure*}

In Fig.~\ref{profiles_n7589}, we demonstrate the results of our data analysis for \N7589. The designations are the same as in Fig.~\ref{profiles_malin2}. As one can see in Fig.~\ref{profiles_n7589}, the kinematics of the ionized gas of \N7589 is very complex: The velocity dispersion measured from the line [O\iii] rises with the distance from the centre, which is not seen in H$\beta$. 
It leads to a lower velocity and a shallower velocity gradient for [O\iii] in comparison to those derived from H$\beta$. 
This likely happens due to the presence of an AGN in \N7589 (see Sect.~\ref{gen_prop}). 
The stellar age appears to be younger for the innermost part in comparison to that at larger radii. 
It and the non-zero values of $h3$ and $h4$ could be an indirect indication for the nuclear disc in the galaxy. 

\begin{figure*}
\includegraphics[width=0.9\linewidth]{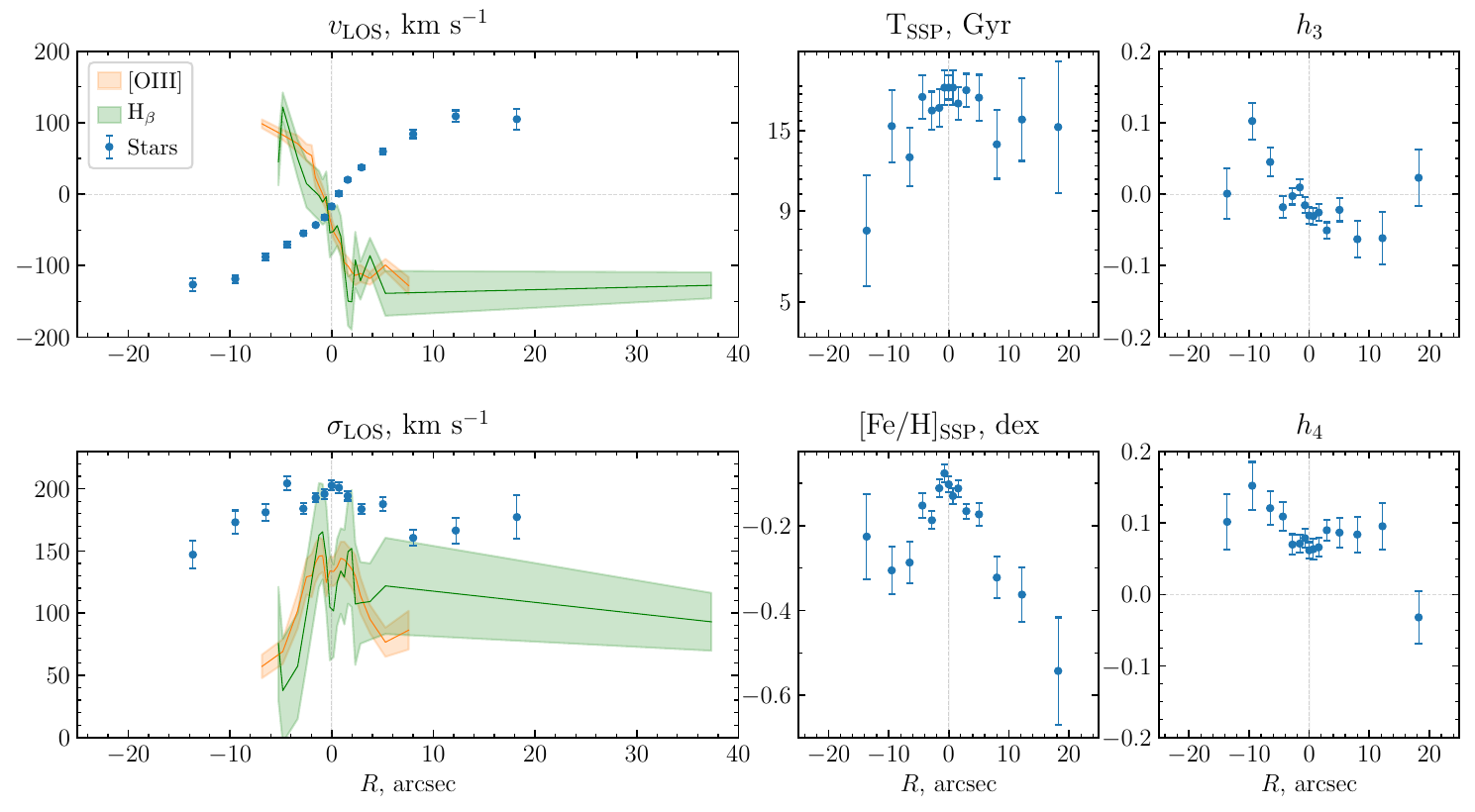}
\caption{Analysis of spectral observations of \U1382. The designations are the same as in Fig.~\ref{profiles_malin2}.}
\label{profiles_u1382}
\end{figure*}

In Fig.~\ref{profiles_u1382}, we present the results for \U1382. 
The  gaseous component counter rotates with respect to the stars at all radii. 
The ionized gas is corotating with the extended \HI disc if we compare our profile with the \HI velocity map presented in \citet{Hagen2016}. 
The stars in the centre are very old and show the considerable increase of metallicity. 
The age of stellar population of HSB part of the galaxy that we obtained in the current paper is significantly higher than that derived by \citet{Hagen2016} from SED fitting that is likely caused by the presence of a small fraction of young stars that strongly affect the blue part of the SED. 
The $h_3$ value is close to zero, at the same time $h_4$ is roughly 0.1 which could indicate the presence of a kinematically decoupled stellar component. 

\begin{figure*}
\includegraphics[width=0.9\linewidth]{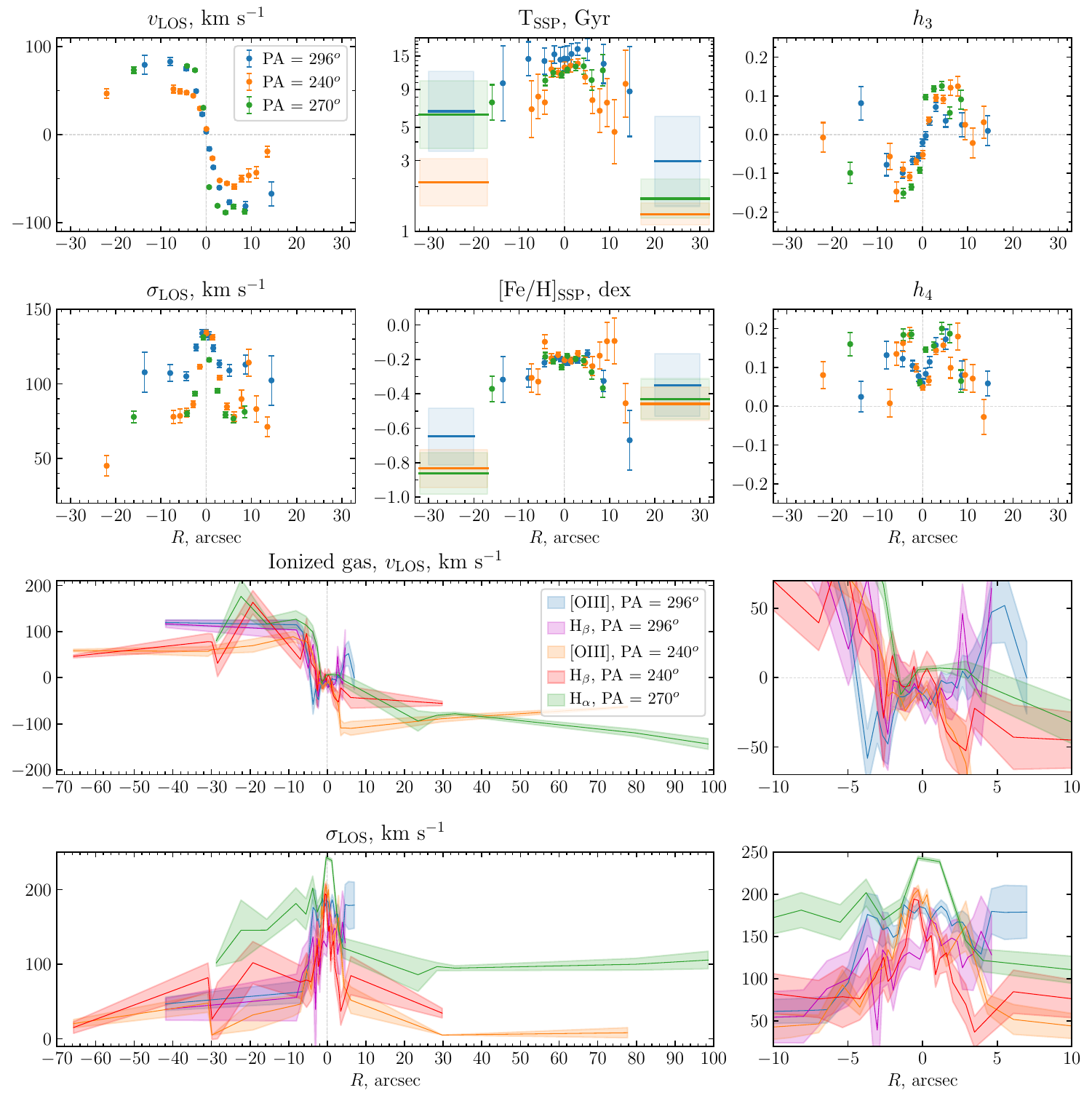}
\caption{Analysis of spectral observations of \U6614 for three different position angles obtained at BTA and Gemini. Top two rows, left-hand panel: stellar kinematics, $v$ (top panel) and $\sigma$ (bottom panel); centre panel: stellar populations, age (top panel) and metallicity (bottom panel); right-hand panel: LOSVD deviation from the Gaussian shape, $h3$ (top) and $h4$ (bottom panel). Two bottom rows: $v$ (top panel) and $\sigma$ (bottom panel) of ionized gas for different PA and emission lines (see the legend). Left- and right-hand panels demonstrate full radial profiles and their zoomed central regions correspondingly.}
\label{profiles_u6614}
\end{figure*}

Fig.~\ref{profiles_u6614} shows the kinematics and stellar population profiles for GMOS-N and SCORPIO observations of \U6614. 
The kinematics in the inner region of \U6614 is also very complex. The motion of the ionized gas is different from that of the stellar population. In the innermost region ($R\leq5$~arcsec), the ionized gas does not rotate or even shows signs of counterrotation. In contrast to \UGC1922, the inverted velocities are visible here only for one position angle $PA=296^{\circ}$ and are concentrated in the innermost region unlike in \UGC1382 where the counterrotation of the gas is global. Hence, probably this feature traces the AGN-driven gas outflow noticed by \citet{Ramya2011} rather than a kinematically decoupled nuclear component. However, a high-resolution two-dimensional velocity field is needed to draw firm conclusion. 
The velocity dispersion measured from the [O\iii{}] line differs from that of H$\beta$, similarly to \N7589, where the AGN also can affect the kinematics.

Another peculiarity of the ionized gas LOS velocity profile is a drop of the velocity on both sides from the centre at the radius of $\sim$30~arcsec, in the vicinity of the inner bright ring of the galaxy. This drop manifests itself also in the velocity profile of stars.  The change of the gradient of the velocity is similar to that often observed in galaxies with bars \citep[see, e.g][and references therein]{Saburovaetal2017}. It is possible that the ring  we see in UGC~6614 is a resonant ring. It could be oval, elliptical ring in which elliptical streamlines occur and non-circular motions show themselves in the decrease of the LOS velocity. Possibly, there was a weak bar in UGC~6614 before, which aided in accumulation of the material in the ring.

\citet{McGaugh2001} also obtained ionized gas LOS velocity profile of \UGC6614 in the H$\alpha$ line from the long-slit observations at $PA=108^{\circ}$. In the inner region of their profiles, one can see the hint of the change of the velocity gradient. The rotation curve obtained in current paper is in a good agreement with their data -- showing rotation velocity amplitude of $\sim 200$ \kms and the minimum at 10~kpc (see Fig. \ref{profiles_u6614}).

The stellar population of the bulge of \U6614 is old and metal-rich. The age of stars decreases with the distance from the centre and appears to be about 2--3 Gyr outside the region where the bulge dominates the luminosity. The metallicity of stars in this radius is roughly $-0.5$~dex.
The values of $h_3$ and $h_4$ are non-zero, in particular $h_3$ anticorrelates with the velocity and $h_4$ has a minimum in centre, which is usually observed in barred galaxies 
\citep[see, e.g.][and references therein]{Saburovaetal2017}.


\subsection{Surface photometry of \N7589 and \UGC6614}
To estimate the structural parameters of \N7589 and \UGC6614, we performed their surface photometry. We took publicly available data from the Subaru HyperSuprimeCam Strategic Survey DR2 for \N7589 and Zwicky Transient Facility (ZTF) \citep{2019PASP..131a8002B} survey for \U6614, both in the {\it r} band. We performed the isophote analysis using the {\sc ellipse} task in the {\sc PHOTUTILS python} library. Then we found a multicomponent best-fitting model by minimizing $\chi^2$  statistics for several light profiles. During this procedure, we convolve our model with a PSF, which we derived by fitting unsaturated stars in the same image.

\N7589 has a complex morphology in the centre which includes two bars and a ring. We excluded the central 17~arcsec from our analysis and fitted only the outer part of the light profile by a Sersic component and an exponential disc. \revtwo{As we show below, the masking of the central part of the profile has little effect on the DM halo parameters (that is the main goal of the approach).}

\U6614  spans two adjacent fields in ZTF, hence we co-added them into one mosaic using the {\sc swarp} software \citep{2010ascl.soft10068B}. Unlike ZTF-images, images in other surveys like SDSS or Legacy Survey, where spatial resolution is significantly better, have artefacts near bright objects which are usually caused by the local sky background subtraction algorithm. In case of \UGC6614 these artefacts prevent a precise light profile decomposition that includes an LSB disc, that is why we decided to use much shallower ZTF data unaffected by sky oversubtraction. We fitted the profile of \U6614 by two Sersic components. A one-dimensional multicomponent light profile decomposition is known to be unstable with respect to the additive background in the original image and also requires a rather fine-tuned initial guess \citep[see e.g.][]{Chilingarian2009}, therefore it is crucial to use data with a precisely subtracted sky background.

\begin{figure}
\includegraphics[width=0.8\linewidth,trim={0cm 0cm 1cm 1cm}]{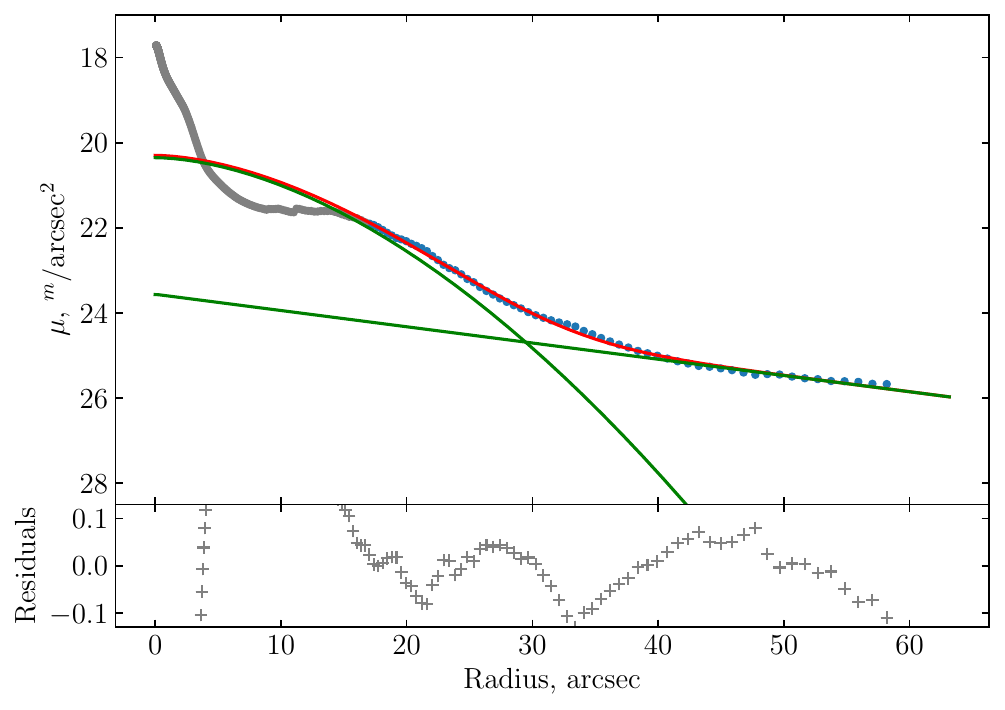}
\caption{The results of the decomposition of the HSC {\it r}-band surface brightness profile of \NGC7589 into Sersic and exponential disc components. The central part inside the 17-arcsec ring is excluded from the analysis and greyed out.\label{NGC7589_sb}}
\end{figure}

\begin{figure}
\includegraphics[width=0.8\linewidth,trim={0cm 0cm 1cm 1cm}]{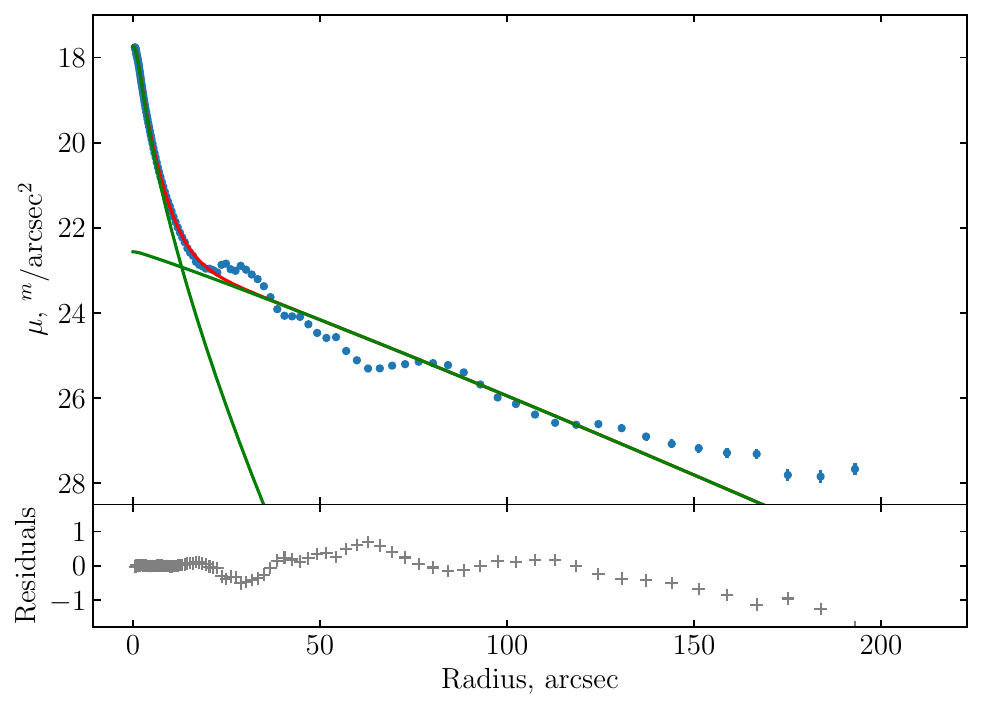}
\caption{The results of the decomposition of the {\it r}-band surface brightness profile of \UGC6614 into two Sersic components.\label{UGC6614_sb} }
\end{figure}

We demonstrate the results of the decomposition of the profiles into inner and LSB disc components for \N7589 and \U6614 in Figs.~\ref{NGC7589_sb} and~\ref{UGC6614_sb}. The parameters of LSB discs and inner components are provided in Table~\ref{UGC6614_sb_tab} for both galaxies.\footnote{\revtwo{We also attempted to use a model with a Sersic inner component and an outer exponential disc for \UGC6614, but the model with two Sersic components yielded better fitting quality. The deviation of the model profile from the observed data at large radii does not affect the subsequent analysis because (i) we used a non-parametric description of the outer disc component as a difference between the total and bulge profiles in the rotation curve decomposition for \UGC6614 and (ii) a radial velocity profile does not reach the outermost part of the disc where the deviations are significant.}}


\begin{table*}
\begin{center}
\caption{The parameters of the {\it r}-band  decomposition of the surface brightness profiles of \N7589 and \U6614. For \U6614, the disc was modelled by a Sersic profile, for \N7589, we used a pure exponential disc. The columns are (1) galaxy name; (2) total magnitude of the Sersic inner component; (3) effective radius of the Sersic inner component; (4) Sersic index of the inner component; (5)--(7) the same as (2)--(4) but for the outer disc component.\label{UGC6614_sb_tab}}
\renewcommand{\arraystretch}{1.0}

\begin{tabular}{cccccccc}
\hline
\hline
Galaxy &$m_{in}$ & $(R_{e})_{in}$ & $n_{in}$& $m_d$ & $(R_{e})_d$& $n_d$\\
&mag & arcsec & & mag & arcsec &   \\

\hline
\N7589&$13.41 \pm 0.05$  & $12.03 \pm 0.17$ & $0.57 \pm 0.02$ & $14.33 \pm 0.09$ & $47.74 \pm 2.23$ &1\\

 \U6614&$13.88 \pm 0.03$  & $4.35 \pm 0.08$ & $1.52 \pm 0.02$ & $13.03 \pm 0.08$ & $51.63 \pm 2.20$ & $0.93 \pm 0.11$ \\
\hline

\end{tabular}
\end{center}
\end{table*}

\section{Dynamical modelling using the rotation curve decomposition}\label{massmod}

The dark matter halo is known to play an important role in galaxy evolution. 
Here, we derive the parameters of dark matter haloes in gLSBGs to compare them to HSB galaxies. In the previous studies of gLSBGs, \cite{Pickering1997, deBlok2001, Hagen2016, Lelli2010, Junais2020, dipaolo} also performed mass modelling using different profiles of dark matter halo (Einasto, NFW, Burkert and pseudo-isothermal) for different galaxies, hence no direct comparison of the objects is possible. Therefore, here we perform the rotation curve decomposition using the same dark matter halo profile for all galaxies in our sample. 
We combine the optical data presented in this paper with the published \HI data, and then decompose the rotation curves for \Malin2, \N7589, \U1382, and \U6614. 
We derived the optical rotation curves from the H$\beta$ and [O\iii{}] emission lines corrected for the systemic velocity, symmetrically reflected about the galaxy centre and de-projected using the inclination angle provided in Table~\ref{tabproperties}.\footnote{\revone{For \UGC6614 three cuts are not along major axis, which was taken into account, see below.}}

In the decomposition procedure, we included the following components: a stellar disc, a bulge, an \HI disc, and a \citet{Burkert1995} dark matter halo.\footnote{We chose Burkert profile to compare with the sample of giant HSB galaxies where it was used by \citet{Saburova2018}.}   We used the gas surface densities from published \HI observations.
The details of the procedure of mass-modelling are described in \citet{saburovaetal2016}. \revone{We give the specifications of the mass-modelling and the models for each  galaxy in Section \ref{appendix}.}

We fixed the contributions of stellar components according to the colour and spectral information. 
We also performed a decomposition of the combined rotation curve of \Malin1 using the published data. 
The results of mass-modelling for \U1378  and \U1922 are taken from \citet{ Saburovaetal2018, Saburova2019}. 

\begin{figure*}
\includegraphics[width=0.7\linewidth,trim={2cm 7.5cm 2cm 7.5cm}]{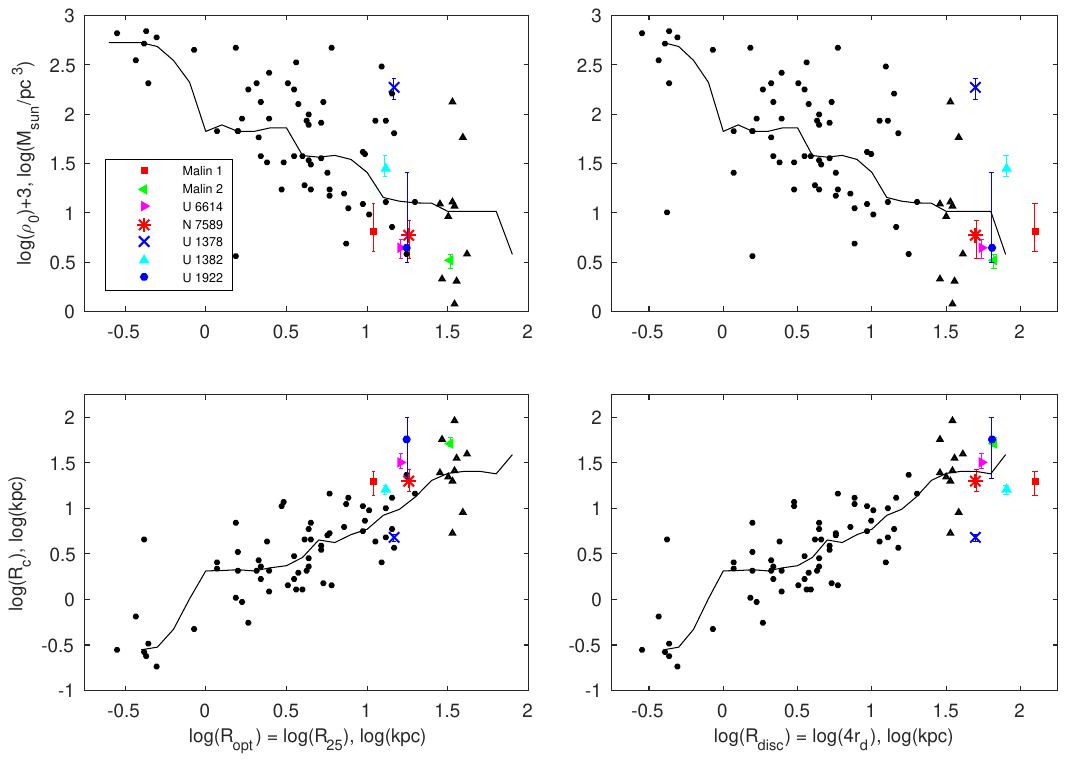}
\caption{Left-hand panel: the optical radius ($R_{25}$) compared to the parameters of dark matter halo, central density (top panel) and radial scale (bottom panel). 
Right-hand panel: the same but for the disc radii from Table~\ref{tabproperties} for gLSB and $R_{25}$ for HSB galaxies. 
The gLSBGs are shown by coloured symbols. 
Black triangles correspond to the giant HSB galaxies from \citet{Saburova2018}, black dots are for galaxies of moderate size, the solid lines are the running medians for non-LSB galaxies. \label{dm_params}}
\end{figure*}

In Fig.~\ref{dm_params}, we plot the parameters of dark matter haloes against the $25$~mag $B$-band isophote of gLSBGs and the sizes of gLSB discs from Table~\ref{tabproperties}. 
We also plot the parameters of intermediate-size and giant HSB galaxies from \citet{Saburova2018} (black dots and triangles). 
The solid lines show the running median for HSB galaxies. 
From  Fig.~\ref{dm_params}, it is evident that gLSBGs behave differently from HSB galaxies, some of them lie on the continuation of the best-fitting relation for HSB galaxies, but some deviate from it. 
This could suggest different nature and formation scenarios for different gLSBGs in our sample.

For \N7589, \Malin2, \U6614, and \U1922\footnote{It is worth noting, the uncertainty in the halo parameters does not draw the clear conclusion in the \revone{cases of \N7589} and \U1922.}, the parameters of the dark haloes agree better with the radius of the gLSB disc than with $R_{25}$. 
The three remaining gLSBGs lie close to HSB galaxies in the diagrams for optical radius. 
This could indicate that the parameters of dark haloes of some of gLSBGs are connected to the sizes of their HSB parts, which can favour the two-stage formation of these systems.
At first, the HSB part is formed, then a build-up of an extended LSB part occurs. 
Interestingly, the HSB-part of \Malin2 is also very extended.

\begin{table*}
\begin{center}
\caption{The parameters of the main structural components of the galaxies with $1\sigma$ uncertainties. The columns contain the following data:
(1) galaxy name;
(2) and (3)~-- radial scale and central density of the DM halo;
(4) mass of the DM halo inside the LSB disc radius given in Table \ref{tabproperties};
(5) central surface density of the bulge; and
(6) disc mass\label{par}.}
\renewcommand{\arraystretch}{1.5}
\begin{tabular}{lrlrl  lrcc}
\hline
\hline

Galaxy	&	\multicolumn{2}{c}{$R_s$}&	\multicolumn{2}{c}{$\rho_0$ }&		\multicolumn{2}{c}{$M_{\rm halo}$}	 &		\multicolumn{1}{c}{$(I_0)_b$}&\multicolumn{1}{c}{$M_{\rm disc}$}\\
&\multicolumn{2}{c}{kpc}&\multicolumn{2}{c}{$10^{-3}$ M$_{\odot}/$pc$^3$}&\multicolumn{2}{c}{$10^{10}$ M$_{\odot}$}&\multicolumn{1}{c}{$10^{3}$ M$_{\odot}/$pc$^2$}&\multicolumn{1}{c}{$10^{10}$ M$_{\odot}$}\\
\hline

\Malin1 &19.6& $^{+     6.4}_{-     5.8} $  &  6.6&  $^{+     5.7}_{-      2.5} $  & 78.3& $^{+     13.2}_{-      15.1} $ &25.2&13 \\
\Malin2 &52.2&$^{+     8.2}_{-     4.8} $&3.3&$^{+     0.5 }_{-     0.6 } $&164.4&$^{+     3.8}_{-      4.1} $&6.4&31\\
\NGC7589 &19.6&$^{+     7.6}_{-    4.1} $&5.9&$^{+     2.4}_{-     2.4} $&34.3&$^{+     4.3 }_{-     3.9} $&-- &1.2\\
\UGC1382 &16.3&$^{+    1.6}_{-    2.1} $&28.6&$^{+    9.3}_{-    5.3} $&157.2&$^{+    4.1}_{-    8.7} $&9.2&5.7\\
\UGC6614 &32.5&$^{+    7.7}_{-   4.6 } $&4.5&$^{+    1.0}_{-   1.0 } $&59.3&$^{+    6.4}_{-   5.0  } $&13.0 &1.5\\
\hline
\end{tabular}
\end{center}
\end{table*}

\subsection{Details on the fitting of rotation curves for individual galaxies}\label{appendix}
\begin{figure}
\includegraphics[width=0.8\linewidth,trim={0cm 0cm 13cm 16cm}]{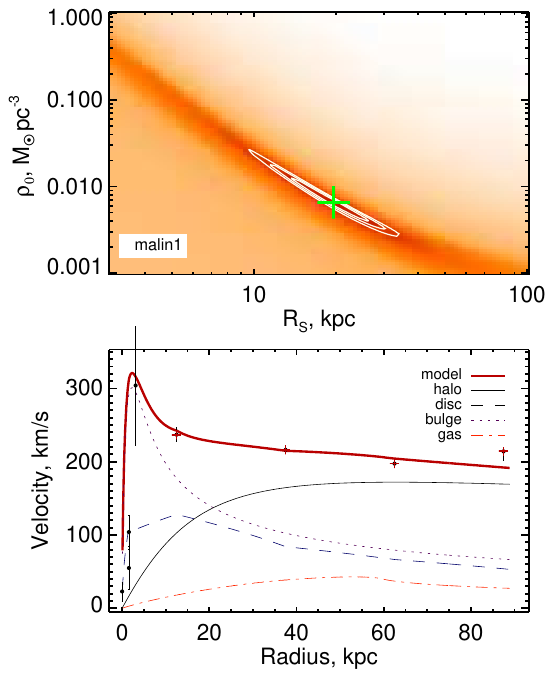}
\caption{The results of the rotation curve decomposition for \Malin1: $\chi^2$ map for the parameters of Burkert dark matter halo (top panel) and the fit of the rotation curve (bottom panel). Black dots on the rotation curve   correspond to the data from \citet{Junais2020} taken along major axis. Red symbols show \HI data. }
\label{massmod_malin1}
\end{figure}

For \Malin1, we utilized the combined optical and \HI rotation curve and surface density profile plus the $R$-band surface brightness profile from \citet{Lelli2010}. We took optical the major axis rotation curve from \citet{Junais2020}.\footnote{We decided to use only the major axis since it was taken with the narrowest slit to avoid possible biases due to a non-homogeneous illumination of the slit and also to minimize the uncertainty because of poorly known orientation of the disc.}
We used a non-parametric definition of the contribution of the disc to the rotation curve. 
For this, we obtained the bulge parameters from the fitting of the light brightness using a Levenberg--Marquardt non-linear least-squares minimization routine in  \textsc{idl} \citep{Chilingarian2009}. 
We fix $R$-band mass-to-light ratios to 3.0 for the disc and limited it to the range 2,$\dots$,6 for the bulge, which correspond to the densities in agreement with those found by \citet{Boissier2016}. 
We present the results of the rotation curve fitting in Fig.~\ref{massmod_malin1}. 

Our estimate of the dark halo radial scale appears to be higher than that obtained by \citet{Lelli2010} but within the range given by \citet{Junais2020} even keeping in mind that they used pseudo-isothermal dark matter profiles rather than the Burkert profile.

For \Malin2, we constructed a combined ionized gas (current paper) and \HI rotation curve \citep{Pickering1997}. We use the parameters of the bulge obtained in $g$ band from the light profile published in \citet{Kasparova2014} (Sersic index $n=0.89\pm0.00$, central surface brightness $\mu_0=18.47\pm0.06$~ mag\,arcsec$^{-2}$, effective radius $R_e=1.10\pm 0.05$~arcsec), and a non-parametric definition of the surface density profile of the disc. The disc $g$-band mass-to-light ratio was limited in the range 1.98,$\dots$,3.89 according to the SED fitting (lower value) and the criterion of the marginal gravitational stability of the disc applied to the stellar velocity dispersion data at $r=20$~arcsec from \citet{Kasparova2014} in a similar way as it was done for \U1378 \citep[see][]{Saburova2019}.

\begin{figure}

\includegraphics[width=0.8\linewidth,trim={0cm 0cm 13cm 16cm}]{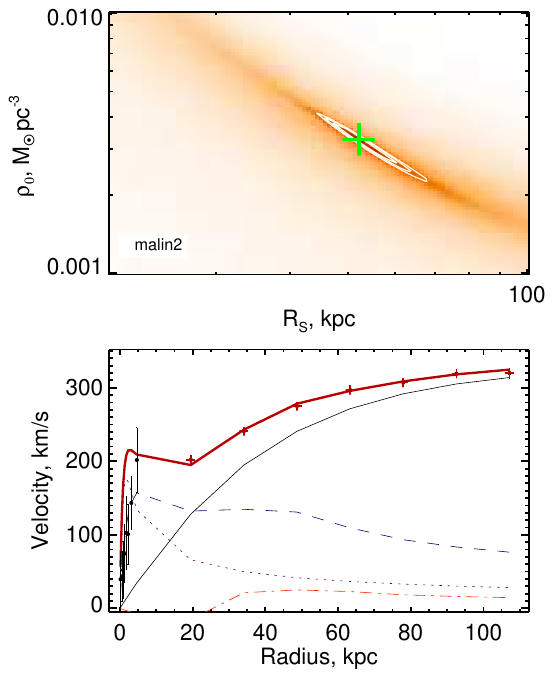}
\caption{The results of the rotation curve decomposition for \Malin2. $\chi^2$ map for the parameters of Burkert dark matter halo (top panel) and the fitting of the rotation curve (bottom panel). Black and red symbols on the rotation curve correspond to optical and \HI data respectively.}

\label{massmod_malin2}
\end{figure}

The value following from the marginal gravitational stability is about twice as high as that resulting from the spectral and SED fitting obtained in \citet{Kasparova2014}, which can indicate the mild overheating of the disc, especially if one takes into account the uncertainty because of the stellar IMF.  
The bulge mass-to-light ratio was limited to 3.33,$\dots$,5.0 according to \citet{Kasparova2014}. 
The results of the decomposition are shown in Fig.~\ref{massmod_malin2}. 
One can see that for \Malin2, the model of its rotation curve rises too steep compared to observations. The probable reason of this discrepancy is the resolution of the spectral data that could smooth the steep rise of the rotation curve in the inner 2~kpc where the inconsistency is observed (the seeing quality corresponded to the resolution 1.3~kpc).

The radial scale of dark matter halo appears to be in good agreement with that found by \citet{Kasparova2014} if one applies the transformation coefficient from a pseudo-isothermal to Burkert dark halo profiles from \citet{Boyarsky2009}. \revone{The core radius found by \citet{dipaolo} is higher than that found in the current paper.}

\begin{figure}
\includegraphics[width=0.8\linewidth,trim={0cm 0cm 13cm 16cm}]{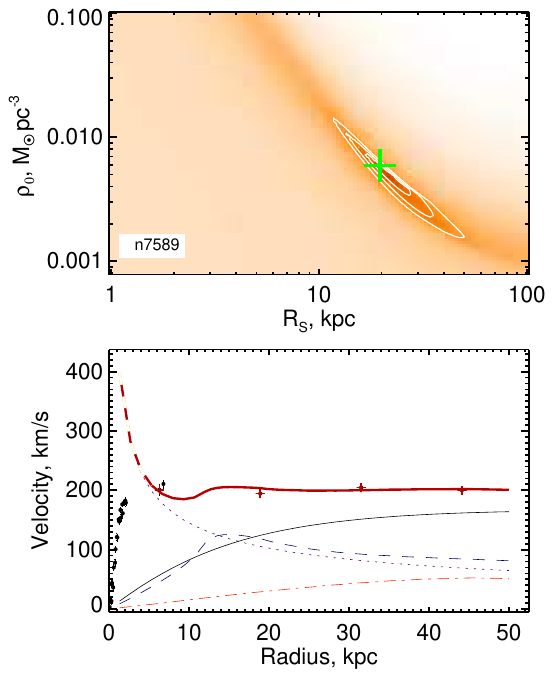}

\caption{The results of the rotation curve decomposition for \N7589. The designations are the same as in Fig.~\ref{massmod_malin2}. Central part of the rotation curve was masked-out during the fitting. }
\label{massmod_n7589}
\end{figure}

For \N7589, we performed the decomposition of the {\it r}-band light profile obtained from the SUBARU HSC data \citep{hsc2019}. The surface brightness profile shows complex behaviour in the inner part due to the presence of  two bars, a nuclear mini-bar with the radius of 2.5~arcsec and the `normal' large-scale bar with the radius of 15~arcsec. Because of this, we calculated the contribution of the inner component to the rotation curve only approximately by considering it as the point mass $V^2=GM/R$ and masked out the inner part of the rotation curve during the fitting. The mass $M$ was estimated from the $r$-band  luminosity in the aperture with the radius of 17~arcsec without the contribution of the LSB disc ($1.7\times10^{10} L_{\odot}$).
The \HI data are taken from \citet{Lelli2010}. The inner rotation curve was derived in this paper from the ionized gas (H$\beta$) ~kinematics. 
The mass-to-light ratio of the inner component was limited in the range $1.4,\dots,2.8$ during the fitting. 
The lower value corresponds to the stellar population with the age of 4~Gyr and solar metallicity (see Fig.~\ref{profiles_n7589}) for the Kroupa IMF. 
The upper value comes from the bulge colour $g-r=0.76$  obtained from the SDSS images and the \citet{Roediger2015} model $M/L$--colour relations. 
For the disc, we considered the $M/L_r$ in the range 1.3$\dots$2 according to its colour $g-r=0.56$ estimated from the SDSS images and model relations from \citet{Roediger2015} and \citet{Bell2003}. 
The result of the rotation curve modelling for \N7589 is shown in Fig.~\ref{massmod_n7589}. \revone{The fact that we masked out the inner part of the rotation curve makes the model results more uncertain. This leads to the uncertainties of dark halo parameters of 50 per cent \citep[see,][]{Saburova2016}.}

The dark halo radial scale is lower than that found by \citet{dipaolo} but is in good agreement with that obtained by \citet{Lelli2010} for a pseudo-isothermal dark halo, however, they maximized the contribution of the stellar component, which we did not do here.

\begin{figure}
\includegraphics[width=0.8\linewidth,trim={0cm 0cm 13cm 16cm}]{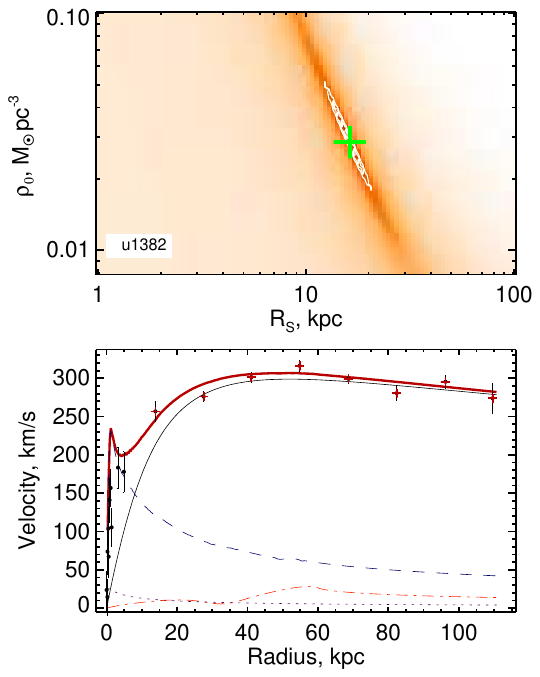}
\caption{The results of the rotation curve decomposition for \U1382. 
The designations are the same as in Fig.~\ref{massmod_malin2}. }
\label{massmod_u1382}
\end{figure}

For \U1382, the ionized gas is co-rotating with the \HI disc, which allowed us to build the combined rotation curve. 
We took the \HI data from \citet{Hagen2016}, the structural parameters of the bulge from \citet{Hagen2016}, and a non-parametric definition of the disc surface density profile -- we obtained it as a difference between the total and a bulge $r$-band surface brightness profiles. 
The $r$-band mass-to-light ratio of the bulge was limited in the range 3,$\dots$,5 according to the stellar metallicity $-0.1$ and old age 13~Gyr for   \citet{Kroupa2001} and  \citet{Salpeter1955} stellar IMF (see Fig. \ref{profiles_u1382}). 
For the disc the limits 1.2,$\dots$,2.66 come from the relation by \citet{Roediger2015} and the $g-r$ colour of the disc and from \citet{Hagen2016}. 
We present the results of modelling in Fig.~\ref{massmod_u1382}. \citet{Hagen2016} also performed the mass modelling of the rotation curve of \UGC1382, however, they used the Einasto and NFW dark halo density profiles. Their model shows a similar behaviour to ours, the baryonic contribution to the rotation curve dominates only in the inner 10~kpc, while the dark halo dominates at larger radii.

For \U6614 due to the complex gas kinematics in the inner region we excluded the innermost points (R$\leq 1.6$~kpc) from our analysis. We calculated the ionized gas rotation curve from the two spectral cuts ($PA=240, 270^\circ$) based on the measurements in H$_{\alpha}$ and [O{\sc iii}] emission lines. We estimated the circular velocity from the observed line-of-sight velocity and the radial coordinate $R$ taking into account inclination of the disc and the angle between the radius-vector of a given point and the major axis of a galaxy $\phi$: \revone{$ V(R) = \frac{V_r(R)\sqrt{(sec^2(i)-\tan^2(i)\cos^2(\phi))}}{\sin(i) \cos(\phi)}$; $ R = R_\phi (\sec^2(i) - \tan^2(i) \cos^2 (\alpha))$.}
\revone{Here,  $ R_\phi $ is the radius in the sky plane, $V_r$ is the line-of-sight velocity corrected for the systemic velocity, $i$ is inclination of the disc.} We adopted the inclination according to \citet{Pickering1997}. The situation with the position angle is more complex since the value $PA=296^\circ$ did not agree with our kinematics in the inner region of \UGC6614, thus we adopted $PA=265^\circ$ for the inner 35~arcsec, and $PA=296^\circ$ for the outer data points.

\begin{figure}
\includegraphics[width=0.8\linewidth,trim={0cm 0cm 13cm 16cm}]{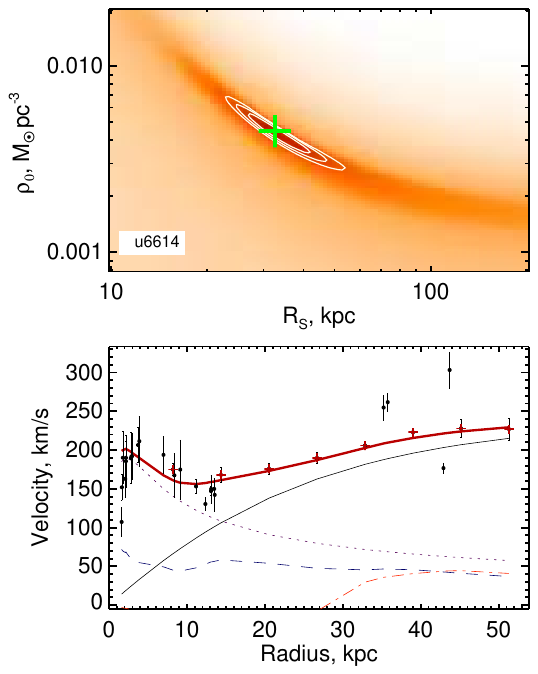}
\caption{The results of the rotation curve decomposition for \U6614. The designations are the same as in Fig.~\ref{massmod_malin2}. }
\label{massmod_u6614}
\end{figure}

We added the ionized gas rotation curve to \HI velocities from \citet{Pickering1997} and obtained the combined rotation curve for the fitting procedure. 
We used the structural parameters of the bulge from our analysis of the ZTF survey $r$-band image discussed above. We used a non-parametric definition of the disc profile as a difference between total and bulge surface brightness profiles.
The $r$-band mass-to-light ratio of the bulge was limited in the range: 2.57,$\dots$,4.54. The upper value corresponds to the stellar population with the age of 12~Gyr and metallicity $-0.2$ for the Salpeter IMF, the lower value is for the age 12~Gyr for the Kroupa IMF. 
The range of disc mass-to-light ratio was estimated from the $(g-r)=0.4$~mag colour of the disc and the $M/L_r$--colour relation from \citet{Roediger2015} taking into account the age 1~Gyr and the metallicity $-0.7$ dex in the region outside the dominance of bulge: 0.4,$\dots$,2.3. We present the fitting results for \U6614 in Fig.~\ref{massmod_u6614}. As one can see from the figure, the model includes the HSB part associated with the pseudobulge component and the LSB part in a form of a non-parametrical disc. This model looks similar to that of \Malin1 and \NGC7589 constructed by \citet{Lelli2010}. 

\citet{deBlok2001} also performed the mass modelling for \UGC6614 using pseudo-isothermal and NFW dark halo density profiles. For the pseudo-isothermal halo, they derived the dark halo radial scale 12.18~kpc using a constant stellar $R$-band mass-to-light ratio 1.4. It appears to be in good agreement with the scale obtained here for the Burkert profile if one applies the transformation coefficients from \citet{Boyarsky2009}. \citet{Pickering1997} obtained somewhat higher dark halo radial scale of 24.48~kpc, but they considered only \HI data as the only available at the moment.

\section{Scaling relations}\label{sr}
\subsection{Baryonic Tully--Fisher relation for gLSBGs}

\begin{figure}
 \includegraphics[height=6.0cm,trim={1.5cm 1.5cm 1cm 1.5cm}]{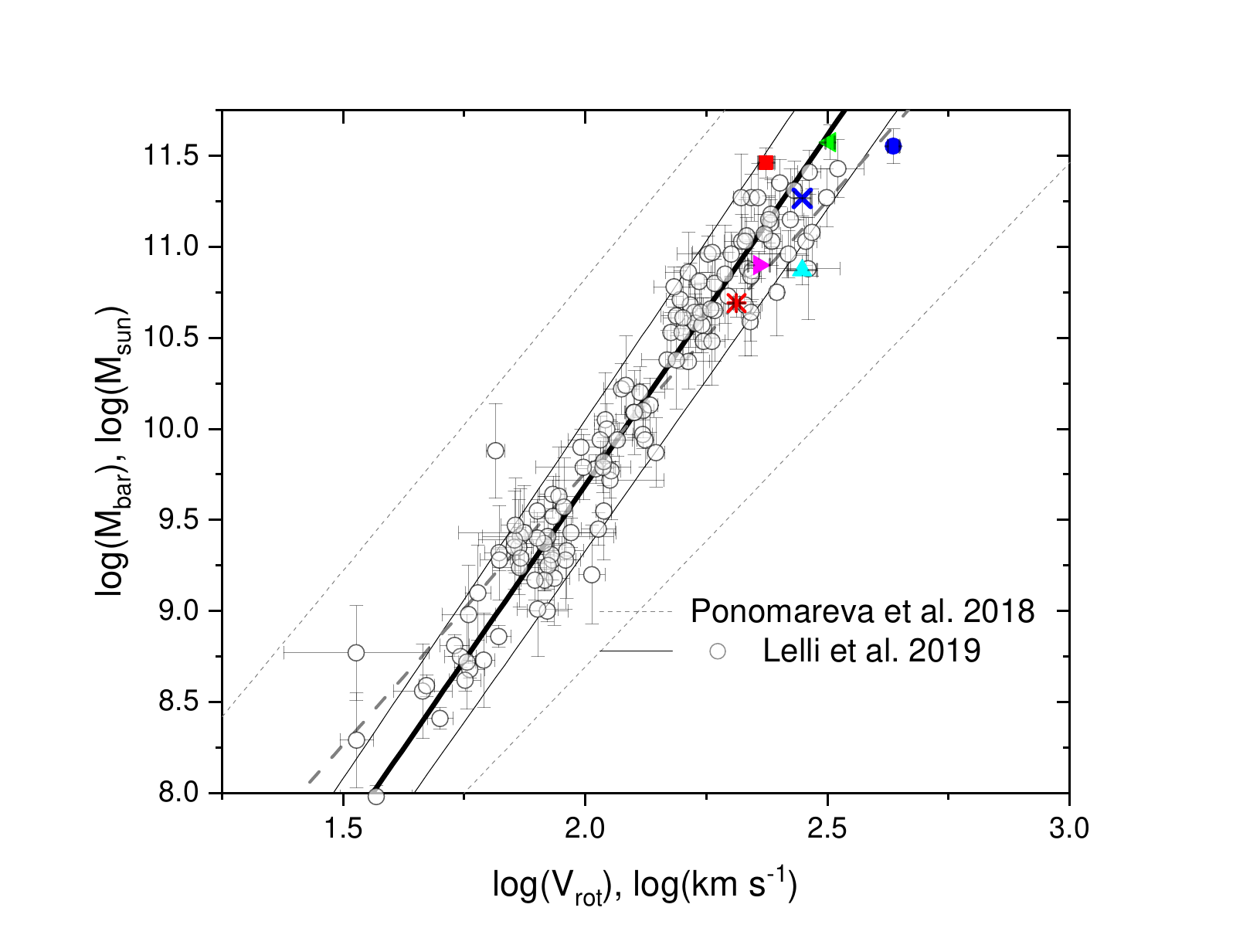}
 \caption{The position of the gLSBGs on the baryonic TF relation. Continuous and dashed lines correspond to the regressions found by \citet{Lellietal2019} and \citet{Ponomareva2018}. Thick lines show the relations and thin lines give the uncertainties of the parameters of the each regression. Open circles demonstrate the sample from \citet{Lellietal2019}. Different symbols denote different gLSBGs as in Fig.~\ref{dm_params}.}
\label{btf}
\end{figure}

The baryonic TF (\citeyear{1977A&A....54..661T}) relation \citep{Sprayberry1995,McGaughSchombert2015}  connects the total baryonic mass (gas+stars) to the maximal circular velocity in a galaxy. In Fig.~\ref{btf}, we display seven gLSBGs from our sample compared to the best-fitting linear correlations for the baryonic TF relation from the literature. Continuous thick line \revone{and open circles}  and dashed thick line show the relations obtained by \citet{Lellietal2019} (the one with the lowest scatter) and \citet{Ponomareva2018} (the final one) correspondingly. Thin continuous and dashed lines give the uncertainties of the correlations given in the corresponding papers. Our gLSBGs are shown with the same symbols as in Fig.~\ref{dm_params}. We computed the baryonic mass as a sum of stellar disc and bulge masses from Table~\ref{par}, \citet{Saburova2018, Saburova2019}, and the \HI masses from Table \ref{tabproperties}. For \UGC1922, the disc mass is estimated from the photometric mass-to-light ratio. The velocities and \revone{their errors} are also taken from Table \ref{tabproperties}. \revone{The errors of baryonic mass are calculated in a similar way as in \citet{Lelli2016} and include the errors of gas mass from Table~\ref{tabproperties}, the uncertainty of the mass-to-light ratios of 0.11 dex following \citet{McGaughSchombert2014} and the uncertainty of the luminosity of order of 10 per cent.} 

As one can see in Fig.~\ref{btf}, gLSBGs occupy the top right-hand corner of the TF relation with high rotational velocities and large baryonic masses. Most of them lie within the uncertainty of the correlation found by \citet{Lellietal2019}. 

According to \citet{Ogle2019}, the baryonic TF relation breaks for rotational velocities higher than 340~\kms, however, in our sample only \UGC1922 rotates that fast. This galaxy has the largest deviation from the relation found by \citet{Lellietal2019} at the same time not being an outlier from the regression found by \citet{Ponomareva2018}. 

\subsection{Star formation in gLSBGs}

Numerous efforts were made by different teams to probe the star formation process in the low-density regime \citep[see, e.g.][]{ Bigiel2010, zasov2012, Bacchini2020} because different astrophysical phenomena might govern star formation in low- and high-density regions. The extended discs of gLSBGs represent a particularly interesting case of a low-density environment because they often have very substantial global SFRs distributed over an enormous area. 

\begin{figure}
 \includegraphics[height=6.0cm,trim={1.5cm 1.5cm 1cm 1.5cm}]{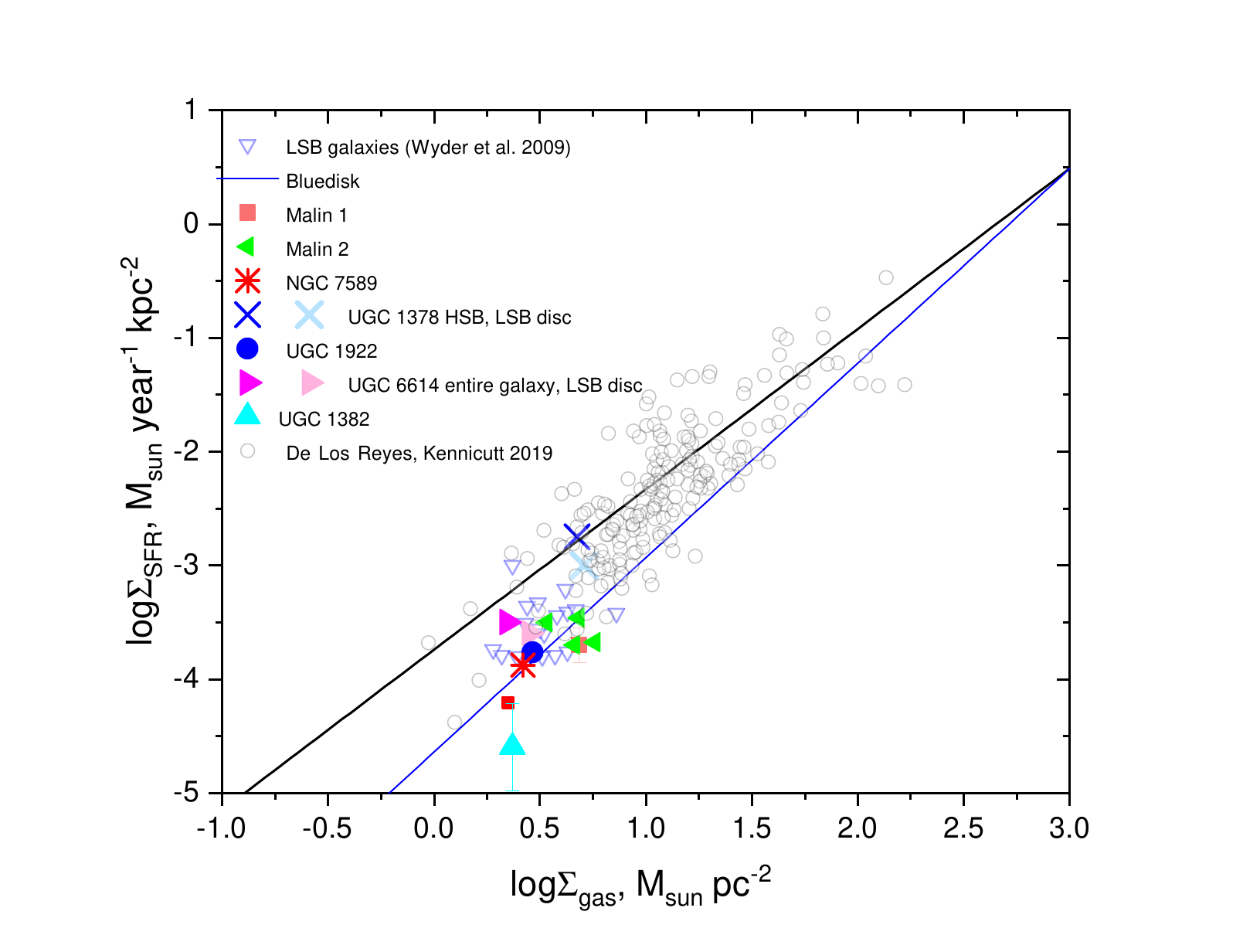}
 \caption{The SFR surface density vs gas surface density diagram. The position of each gLSBG is demonstrated by different symbols (see the legend). \revone{The black line and faint open circles correspond to the best-fitting relation by \citet{delosReyes2019}.}  The blue line shows the best-fitting relation for the nearby galaxies with unusually high \HI mass fractions from the `Bluedisk' project \citep{Roychowdhury2015}.  Faint triangles show LSB galaxies from \citet{Wyder2009}.}
\label{sfr}
\end{figure}

We display the \revone{integrated} SFR surface densities $\Sigma_{\mathrm{SFR}}$ for the seven gLSBGs from our sample on the Schmidt--Kennicutt relation connecting $\Sigma_{\mathrm{SFR}}$ with the gas surface density in Fig.~\ref{sfr}. For all gLSBGs except \Malin2, we plot the \HI surface density instead of the total gas surface density because the reliable spatially resolved molecular gas measurements were not available for them \revone{and it is important that molecular gas density measurement are made for the same areas as the densities \HI and SFR. As we also noted  above the upper estimates of the molecular gas mass found for \NGC7589 and \UGC6614 are roughly 10 and 100 times lower than that of \HI masses, so the inclusion of molecular gas would not change the position of the gLSBs on the Schmidt--Kennicutt diagram significantly}. \revone{For \Malin2, we give the estimates integrated for the bins with available molecular gas measurement. }The $\Sigma_{\mathrm{SFR}}$ value was taken from \citet{Kasparova2014} and the gas surface density from \citet{Das2010}.   Different symbols show the gLSBGs from our sample as indicated in the legend. \revone{For \Malin1,  we took the integrated $\Sigma_{\mathrm{SFR}}$ and \HI surface density from \citet{Wyder2009} and for the comparison we plot also the local value by fainter colour:} $\Sigma_{\mathrm{SFR}}$ at $R=26$ kpc from the centre taken from \citet{Junais2020} and the \HI surface density at the same galactocentric distance from \citet{Lelli2010}. For \NGC7589, \UGC1922, and \UGC6614, we calculated the $\Sigma _{\mathrm{SFR}}$  from the GALEX FUV data and plotted it against the mean \HI surface density within the same radii as the UV-based SFR. The data sources for \HI densities are given in Table~\ref{tabproperties}. For \UGC 1378,  we took the estimates from \citet{Saburova2019}. For \UGC1382, we estimated the average values of the SFR and \HI surface densities using the data from  \citet{Hagen2016}, the former value was also corrected to include helium. 

A thick black line shows a power law with the index of \revone{1.41 found for spiral galaxies by  \citet{delosReyes2019} in agreement with the classical work by \citet{Kennicutt1998}. Faint open circles show the sample from \citet{delosReyes2019}}. Faint triangles denote LSB galaxies by \citet{Wyder2009}.  To compare gLSBGs with extremely H{\sc i}-rich galaxies we also demonstrate by a blue line the best-fitting relation for the \HI surface density of the galaxies from the `Bluedisk' project \citep{Roychowdhury2015}.

As one can see in Fig.~\ref{sfr}, all gLSBGs \revone{besides} \UGC1378 are the outliers from the  \citet{delosReyes2019} relation for HSB galaxies -- they have too low $\Sigma_{\mathrm{SFR}}$ for the given gas surface densities. Their position is in a good agreement with the position of other low surface brightness galaxies found by \citet{Wyder2009}. There is an indication that the star formation efficiency might be connected to the mechanical energy of collisions of giant molecular clouds in HSB discs due to differential rotation \citep{2020MNRAS.496.5211A}. However, this correlation does not hold at low gas surface densities at the same values where the local Schmidt--Kennicutt relation breaks, suggesting a different physical mechanism regulating star formation there. For example, fresh gas accretion from a filament or a gas-rich satellite will lead to the apparent decrease in the star formation efficiency. The oxygen abundance estimates for \UGC1922 disc are quite high ($12+\log(O/H)\approx 8.6$), which argues against the accretion of metal-poor gas at least in this particular case. \revone{As it was demonstrated by \citet{Pickering1997} the \HI surface density is below the critical density for star formation estimated using the dynamical criterion proposed by \citet{Kennicutt1989} in all radii for \Malin2 and \NGC7589 and is close to the threshold value or slightly above it for \UGC6614 and \Malin1 correspondingly, it could be the reason of low values of SFR in these galaxies. At the same time \citet{Pickering1999} studied another giant discy galaxy with  larger amount of star formation occurring within the disc and the large part of the disc possessing the gas density above the critical value, thus validating this criterion.  } The low efficiency of star formation could be related to significantly lower gas volume densities. As shown by \citet{Zasov2011}, the gas volume densities of LSB galaxies are two orders of magnitude lower than those in galaxies with normal surface brightnesses. \revone{\citet{Abramova2008} and later \citet{Bacchini2019, Yim2020} found tight correlation between the volume densities of gas and SFR  which can indicate that taking into account the thickness of the disc could eliminate the break of the relation on the low surface densities. And indeed, as it was shown by \citet{Bacchini2020}, the relation for volume densities holds unbroken for a wide range of volume densities of gas and SFR including low-density regime. It could indicate that the gaseous discs of gLSBs have the high thickness which leads to their outlying position on the gas surface density vs SFR surface density diagram. Another effect that could be important in low-density regime and could lead to the break of the gas surface density vs SFR surface density relation is the influence of the diffuse background both in SFR traces and in atomic gas which was not taken into account in the current paper \citep{Kumari2020}.  }

\section{Discussion}\label{Discussion}\label{form_sc}

\subsection{Formation scenarios of gLSBGs}

The main goal of our study is to choose realistic formation scenarios of gLSBGs based on the observational data of all stellar systems of this type studied in detail up-to-date.
We are trying to understand whether the processes leading to the formation of such unusual galaxies are extremely uncommon making gLSBGs `unique'.

Our observational data suggest that no single formation mechanism can explain all seven gLSBGs presented here. At the same time, we consider only three of the considered scenarios because each of them is consistent with all observations of at least one gLSBG: (i) a scenario involving a major merger, similar to that proposed in \citet{Saburovaetal2018} and \citet{Zhuetal2018}; (ii) an explanation of unusual properties of giant LSB discs through the peculiar properties of the dark matter halo, namely the low central density and the high radial scale of the halo \citep{Kasparova2014}; (iii) a two-stage formation scenario in which the giant disc is formed by accretion of gas on a preexisting `normal' HSB galaxy \citep{Saburova2019}.

In Table~\ref{tab_scenarios}, we give a short summary of the signatures that we expect in each for the three considered scenarios. Table~\ref{gal_scenarios} assesses all possibilities for each galaxy in our sample. 

\begin{table*} 
\caption{Expected signatures of the proposed formation scenarios of gLSBGs. \label{tab_scenarios}
}
\begin{center}
  \begin{tabular}{|l|c|c|c|}\hline 
\diagbox[width=20em]{Signatures}{scenario}&Major merger&Sparse dark halo  & \begin{tabular}{c}
         Gas accretion\\ on a preformed galaxy\\
       \end{tabular} \\
\hline 
DM parameters agree with the giant disc radius&No &Yes&No\\
\hline
Two discs with different scale lengths&No&No&Yes\\
\hline
Dynamical overheating of the disc&Expected&Not expected&Not expected\\
\hline
Disturbed morphology&Expected& Not expected&Not expected\\
\hline
Presence of satellites&Expected&Not expected&Not expected\\

\hline

\end{tabular}
\end{center}

\end{table*}

\begin{table} 
\caption{Assessment on the proposed formation scenarios for each galaxy in our sample.
\label{gal_scenarios}}
\begin{center}
  \begin{tabular}{lccc}\hline \hline 
 &Major merger&\begin{tabular}{c}Sparse\\ Dark halo\end{tabular}&\begin{tabular}{c}
         Gas accretion \\ on a preformed\\ galaxy\\
       \end{tabular}\\
\hline
\Malin1&Possibly&Uncertain&Possibly\\
\Malin2&No&Yes&No\\
\N7589&No&Yes&Yes\\
\U1378&No&No&Yes\\
\U1382&Possibly&No&Possibly\\
\U1922&Yes&Possibly&No\\
\U6614&No&Yes&Possibly\\
\hline
\end{tabular}
\end{center}

\end{table}

We give the detailed discussion for the considered formation scenarios in the next subsections. But first, we explain why we do not find enough of the supporting evidence for some other mechanisms discussed in the literature.

We believe that the scenario of the outer disc formation as the result of a dwarf satellite merger (proposed by \citet{Penarrubia2006} for Messier~31) has difficulties explaining the gLSB disc formation for several reasons. (i) An extended LSB disc often contains a substantial fraction of the baryonic matter in a gLSBG: their masses in \UGC1378 and \UGC1922 are $8\times10^{10}$ and $1.8\times10^{10}$~{\Ms} \citep{ Saburova2018, Saburova2019} (see also Table \ref{par} for the masses of the extended discs obtained in this paper). 
In all our cases, the masses of atomic hydrogen exceed $10^{10}$~{\Ms} (Table~\ref{tabproperties}), and it is 10--100 times more than we can expect for a dwarf galaxy satellite \citep[see e.g.][]{Bettoni2003}. 
(ii)~Another reason is that we do not observe a rapid decline in rotation curve in the peripheries of giant LSB discs \citep[see, e.g.,][]{Kasparova2014} expected in the dwarf satellite merger scenario. Random orientation and high eccentricity of the orbits of infalling satellites predicted by some numerical simulations in the Lambda cold dark matter ($\Lambda$CDM) cosmology will likely lead to the loss of angular momentum by both gas and stars and form a dynamically hot stellar halo with the gas sinking to the centre rather than forming a disc. A potential solution exists if numerous satellites infall onto the gLSBG progenitor from a thin vast rotating plane like the ones discovered around the Andromeda galaxy \citep{2013Natur.493...62I} and Centaurus~A \citep{2018Sci...359..534M} suspected to be aligned with the large-scale structure of the Universe \citep{2015MNRAS.452.1052L}. Such planes despite being recognized as one of the `small scale problems' of $\Lambda$CDM \citep{2017ARA&A..55..343B}, are successfully reproduced in numerical simulations \citep[see e.g.][]{2014ApJ...784L...6I,2020ApJ...897...71S}. However, the feasibility of this gLSBG formation scenario still remains in question because most known gLSBGs do not have many satellites observed in their vicinity, while some such satellites would inevitably survive if a massive accretion of a satellite ring/plane occurred in the recent past or is currently happening.

Another scenario explaining the formation of the expanding ring-like giant LSB structure by a bygone head-on collision with a massive intruder proposed by \citet{Mapelli2008} also seems to be unlikely, since none of the expected signatures of this model are found in the deep images and colour profiles of gLSBGs \citep{Kasparova2014, Boissier2016, Hagen2016}. In addition, the progenitor in this model should already be a massive LSB-galaxy which makes it even less realistic and does not really explain the formation of a giant LSB disc.

\subsubsection{The origin of the central regions of gLSBGs}

One possible way to understand the evolution of a galaxy is to explore its inner regions and a nucleus. The structure of the bulge and the mass of the central black hole can help us to better understand the merger history of the galaxy because they are expected to co-evolve \citep[see e.g.][and references therein]{2013ARA&A..51..511K}.

One can notice from Table~\ref{tabproperties} that gLSBGs host AGN in six out of seven considered cases\footnote{The sevenths galaxy -- \UGC1378 may also contain an AGN according to \citet{Schombert1998}.} -- such AGN  occurrence rate is extremely high compared to HSB galaxies \citep{Ho1997}. The high frequency of low-luminosity AGN signatures in gLSBGs compared to other late-type galaxies was also noticed by \citet{Schombert1998}. 

\begin{figure}
 \includegraphics[height=6.0cm,trim={1.5cm 1.5cm 1cm 1.5cm}]{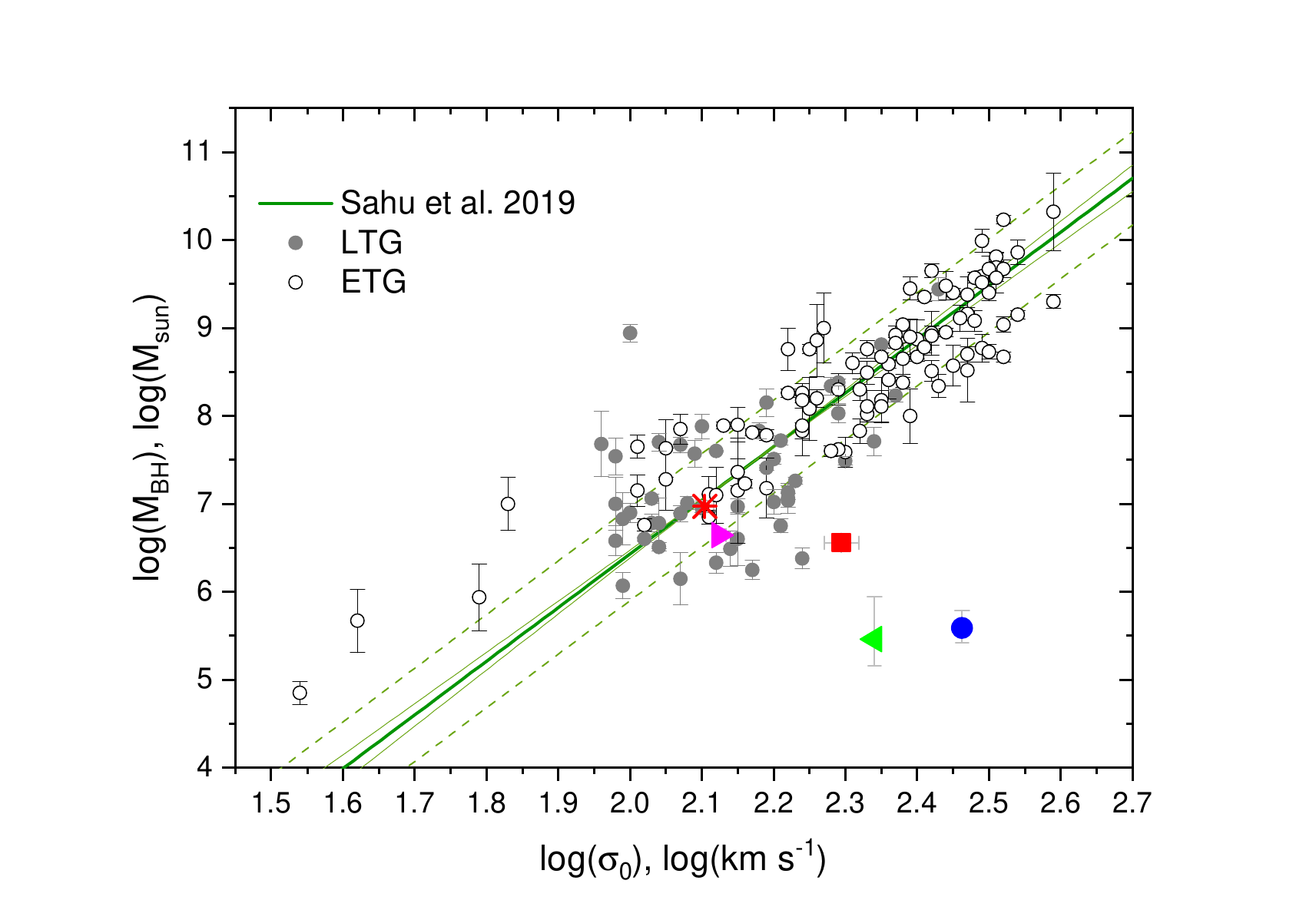}
 \caption{The gLSB galaxies with available measurements of central black hole masses on the mass of the black hole versus central velocity dispersion ($M_{BH}-\sigma_*$) diagram. Open and filled circles show early- and late-type galaxies from \citet{Sahu2019}. Continuous thick line shows the best-fitting relation found by \citet{Sahu2019} for the combined sample of early- and late-type galaxies. Thin lines show the $1\sigma$ uncertainty of the slope and the intercept of the relation. Dashed lines represent the $\pm1\sigma$ scatter in their data. Different symbols depict different gLSBGs in the same manner as in Fig. \ref{dm_params}. The clear outliers are \Malin1, \Malin2, and \U1922.}
\label{mbhsigma}
\end{figure}

Another detail that becomes evident when looking at Fig.~\ref{mbhsigma} and Table~\ref{tabproperties} is that gLSBGs tend to have low masses of central black holes ($M_{BH}$) which sometimes even get close to the intermediate-mass black hole regime. \citet{Subramanian2016} noted a systematic offset of LSB galaxies from the $M_{BH}-\sigma_*$ relation where stellar velocity dispersion $\sigma_*$ is averaged within one effective radius. According to \citet{Subramanian2016}, LSB galaxies tend to have lower masses of black holes than what is expected for the values of velocity dispersions in their bulges.

In Fig.~\ref{mbhsigma}, we show the $M_{BH}-\sigma_*$ for  gLSBGs from our sample and compare them to a large sample of `normal' galaxies and the regression found by \citet{Sahu2019}. The stellar velocity dispersion measurements of gLSBGs come from this study and from \citet{Reshetnikov2010, Saburova2018, Saburova2019}. The adopted black hole masses and corresponding data sources are given in Table \ref{tabproperties}. Open and filled circles demonstrate early- and late-type galaxies from \citet{Sahu2019}. We use the regression obtained by \citet{Sahu2019} for both early- and late-type galaxies (the situation does not change significantly if we consider the regression found only for galaxies with discs). From Fig.~\ref{mbhsigma} it is clear that three of the five gLSBGs with measured black hole masses deviate from the relation found by \citet{Sahu2019}, and \UGC6614 lies on the margin of the $1\sigma$ range having a slightly lower $M_{BH}$ than expected for its $\sigma_*$. This can indicate that the bulges of the considered gLSBGs did not co-evolve with the central black holes and either were formed \emph{in situ} via secular evolution or grew via minor mergers, which are not expected to increase the $M_{BH}$ mass because dwarf galaxies do not often host central massive black holes. \citet{Kormendy2011}, who reported the pseudobulge classification for galaxies with dynamically detected black holes, found out that the black hole masses correlate very little at best with pseudobulge properties.

\citet{Graham2014} showed that pseudobulges are difficult to identify and their classification is quite subjective. However, it should be noted that  \Malin1, \Malin2 (for a model with two discs), \UGC6614, \UGC1378, and \UGC1922 have S\'ersic indices $n<2$ and their bulges do not look rounder than the discs which could be the main argument for their classification as pseudobulges. It makes the high frequency of AGN activity among gLSBGs even more interesting, since the AGN signatures in the emission line ratios decreases significantly for the galaxies with pseudobulges \citep{Yesuf2020}. It could indicate the presence of the current gas supply to the centres of gLSBGs leading to the AGN activity and related to the accretion of material either from a filament or from a gas-rich satellite. The counterrotation of gas that we observe in \UGC1382 and \UGC1922 could also be related to the external gas accretion. \citet{Khoperskov2020} showed that the AGN is efficiently triggered by the retrograde gas infall into a galaxy.

At the same time, according to \citet{Kormendy2011}, the growth of central black holes in pseudobulges is driven rather {\it not} by global processes like major merger but  more by local and stochastic processes. In this regime, black holes are not expected to co-evolve with the bulges and hence their masses might not correlate with the properties of bulges -- exactly what we see in Fig.\ref{mbhsigma} for gLSBGs except \NGC7589 (and possibly \U6614). This could imply that there were not a lot of major merger events in the history of most gLSBGs and their baryonic masses were accumulated by some other processes. The study of globular clusters in \UGC6614 using deep HST images also provides evidence that the galaxy accumulated its stellar mass  through sporadic star formation activity and its star formation history lacks dominant starburst events which could be induced by major mergers \citet{Kim2011}. 

Taking into account all these facts, we conclude the observed properties of central parts and central black holes of gLSBGs support the scenario of the gLSB disc formation by external cold gas accretion either from cosmic filaments or from gas-rich satellites and disfavour their formation by major mergers.

\subsubsection{Gas accretion on a preexisting galaxy (HSB+LSB)}

About a decade ago \citep{Lelli2010} it became clear that some gLSBGs have a complex structure consisting of an inner `normal' HSB early-type spiral galaxy immersed in an extended LSB disc. Discs of such HSB+LSB galaxies cannot be described by a single exponential profile.
We expect that the extent of the dark halo in such galaxies is in good agreement with the $R_{25}$ size of the HSB part. 
The relatively low radial scale of the dark matter halo can indicate that the accreted gas could have the angular momentum differing from that of the HSB disc.
In this case, there is a two-stage build-up of the galactic disc and the main question is: where did the additional material (gas and stars?) for the LSB disc formation come from? 
The environment of such galaxies has to, first, have a source for cold gas accretion, and secondly, a possible close interaction should not tear off the LSB periphery of the disc by tidal effects. If the additional material originates from a cosmic filament, then the environment of such a galaxy should be sparse. 
Observational data for four of seven gLSBGs do not contradict this scenario.

In the case of \Malin1 several facts speak in favour of this formation mechanism. 
\Malin1 shows a complex double disc structure including HSB and LSB parts \citep{Lelli2010}. 
The outer part is much bluer than the inner one \citep{Galaz2015}. 
The age of the extended disc is young \citep{Boissier2016}. 
The current burst of star formation in the extended disc could have been induced by a recent minor merger about 1~Gyr ago \citep{Reshetnikov2010}.

\N7589 also demonstrates the HSB+LSB structure \citep{Lelli2010}. The positional angle of the major axis of \N7589 changes with radius evident from both, photometric analysis and the twist of isovelocity lines in the \HI velocity field published by \citet{Pickering1997}. This can indicate the external origin of the outer LSB structure.  
This galaxy also possesses a massive companion with the velocity difference of roughly 100~{\kms} (\N7603 which has disturbed spiral arms and is twize the size of \N7589).  
Both, the extended blue disc of \N7589 and the disturbed appearance of \N7603 could be the traces of past interactions of these two galaxies. 
The elongated thin spiral arms could be possibly short-living and related to the density wave in the disc induced by a recent tidal interaction between these two systems. 
In this case, the interaction `lit up' the giant disc of \NGC7589 which became more visible with the spiral arms and the blue colour of the extended disc because an interaction induced burst of star formation. 

\U1378 was studied in details in \citet{Saburova2019} who proposed a two-step formation scenario in which the gas accretion onto a `typical' HSB spiral galaxy similar to the Milky Way led to the formation of the extended LSB disc.

\U1382 is a lenticular galaxy with an extended LSB disc. 
Our data reveal that the ionized gas in the centre rotates in the opposite sense to the old stellar disc, but it co-rotates with the giant \HI disc described by \citet{Hagen2016}.
That is, there is a counterrotation of the large-scale discs along all distances from the galaxy center.
Assuming a natural evolutionary link between the extended \HI disc and the LSB disc we expect that the stellar LSB disc also counterrotates with respect to the HSB part.
Our absorption-line spectra unfortunately do not reach the LSB part and cannot be used to support or refute this hypothesis. 
However, a counterrotating LSB disc might contribute to the stellar LOSVD even in the HSB part of the galaxy. 
This hypothesis is supported by non-zero values of $h3$ and $h4$ profiles (see right-hand panel Fig.~\ref{profiles_u1382}) suggestive of a non-Gaussian shape of the stellar LOSVD.
To check this idea we recovered stellar LOSVD in a non-parametric shape applying methodology from \citet{Katkov2011_n524, Katkov2016_n448}. 
Unfortunately, we did not found clear signs of the stellar counterrotation like two distinct peaks or long tails in the LOSVD.
We detected a slightly non-Gaussian shape and a peaked structure of the LOSVD slightly varying from one bin to another likely caused by the noise in the data.
This means that the LSB stellar disc is too faint to be detected in the region dominated by the HSB disc. Therefore, only dedicated ultradeep spectroscopy of the external LSB disc might confirm our expectation about counterrotation of the stellar LSB part in this galaxy.

The \HI disc of \U1382 with a rotational velocity of 280~\kms \citep{Hagen2016} has signs of ongoing star formation (as we can see by the presence of UV-bright spirals in GALEX images).
Our spectrum fitting gives very old age of the central part with the trend of the age decreasing with the radius, i.e. a negative age gradient (see Fig.~\ref{profiles_u1382}).
According to \citet{Hagen2016} the age of the LSB spirals is 4~Gyr older than the HSB part. 
Such a difference with our data can be associated with the age estimation methodology (spectral vs. photometric) and the adopted star formation histories (SSP vs. exponentially declining). 
We believe that the counterrotation of the large-scale \HI disc could indicate the gas accretion from a filament \citep{Algorry2014} on a preexisting early-type disc galaxy (probably, S0). 

\subsubsection{Sparse dark halo}

The second scenario is a non-catastrophic formation of a single disc with a large radial scale in a sparse dark matter halo (large radial scale and low central density). 
    In four cases, \Malin2, \N7589, \U6614 and \U1922, the radial scale of the dark halo is in agreement with the size of an extended disc. One should keep in mind the  uncertainties of dark matter halo parameters for \U1922 and \N7589 are too high to make firm statements on the shallowness of their halo. 
 Normal-sized LSB galaxies are probably formed in haloes with low concentrations of a rapidly rotating host as a result of the centrifugal equilibrium \citep{Mo1998, Bullock2001, kimlee2013}, and the extreme cases of such haloes can host gLSBGs. 
The reasons for the formation of a dark halo with such properties can be found in numerical models of protohalo mergers and they can probably be related to the low-density environment \citep{Maccio2007}.

\subsubsection{Major merger}

In the modern galaxy formation framework, massive galaxies with stellar mass $\sim10^{11}$~{\Ms} should have experienced at least one major merger in their lifetime \citep{Rodriguez-Gomez2015}. 
However, such events are more likely to occur in a dense environment. 
It is expected that a galaxy formed in this way will have a hot stellar disc (if the disc survives at all) and, probably, perturbed morphology \citep{Saburovaetal2018}. 
However, all existing data for gLSBGs except \U1922, show the contrary.
The prominent well-organized spiral structure can indirectly evidence that the discs of most known gLSBGs are thin and, consequently, are not significantly overheated (Zasov et al. in preparation).
To confirm the catastrophic scenario, we need to estimate the stellar velocity dispersion in the region of the LSB disc of a gLSBGs. This is a very challenging observational task. The modern spectrographs allow astronomers to probe velocity dispersions down to 10--15~\kms\ at surface brightnesses as low as $\mu_{g} = 25.5$~mag~arcsec$^{-2}$ in UDGs \citep{Chil2019}, which should, in principle, be sufficient for some gLSB discs. However, because of young and potentially metal-poor stellar populations anticipated in gLSBGs, their spectra are expected to have much shallower absorption lines leading to the drastic increase of uncertainties of internal kinematics for the spectra of the same depth \citep{2020PASP..132f4503C}. A viable solution is to find edge-on gLSBGs where the surface brightness is boosted because of the line-of-sight integration through the disc.

\citet{Zhuetal2018} proposed a major merging scenario for \Malin1 which does not contradict to its observed properties including the absence of the gradient of the stellar age in the disc. We also observe two red satellites projected on-to the LSB disc of \Malin1 \citep{Reshetnikov2010, Galaz2015} which could be the survived remnants of larger galaxies that interacted with it. Their structural properties put them into the rare class of compact elliptical galaxies \citep{2009Sci...326.1379C} proven to be formed via tidal stripping of massive projenitors \citep{2015Sci...348..418C}. A low-luminosity compact galaxy is also orbiting \UGC1382. 
One should keep in mind, however, that the study by \citet{Zhuetal2018} does not mention whether they succeded to reproduce the light profile described by a sum of two exponential components, as it was demonstrated for \Malin1 by \citet{Lelli2010}.

In the case of \Malin2 we can exclude the major merger scenario since its disc is only mildly dynamically overheated and \citet{Kasparova2014} observed a steep metallicity gradient for the gas in the disc which is likely to be flattened during mergers \citep[see, e.g. ][and references therein]{Zasov2015}. 

\U1922 described in detail in \citet{Saburovaetal2018} seems to differ from all other other known gLSBGs. 
It has a strongly overheated stellar disc with clumpy irregular spiral arms. 
The ionized gas in the inner region counterrotates with respect to the outer part of the galaxy. 
\citet{Saburovaetal2018} reproduced most of the observed features of the galaxy in the model of an in-plane merger of giant Sa and Sd galaxies. 
Therefore, \U1922 is likely a result of a major merger. 
This assumption is also supported by the fact that this is the only galaxy in our sample  deviating from the TF relation.

\subsection{Comparison with other galaxies with extended LSB and XUV discs}

The demarcation between gLSBGs and `normal' extended LSB galaxies in the parameter space is rather arbitrary. 
We limit this study to the objects disc radii larger than 50~kpc, which we consider as an informal boundary for gLSBGs.
At the same time, LSB galaxies with moderate sizes have similar properties to gLSBGs and might have similar formation scenarios. 
Therefore, it might be useful to compare our sample against extended LSBGs with disc scale lengths somewhat smaller than 50~kpc. 
One prominent case is \N5533 with the LSB disc with the scale length of approximately 9~kpc  \citep{2007MNRAS.376.1480N}. 
Despite being substantially smaller than \U1378, this galaxy has a similar complex structure that contains an `HSB galaxy' embedded in extended LSB disc \citep{Sil'chenko1998}.
Since the two galaxies have similar structural properties even though at different spatial scales, it will be helpful to compare their general characteristics. 
Stellar populations in the HSB part of \N5533 are also similar to those of \U1378.
The HSB stellar disc of \N5533 is metal-poor and has an intermediate age ($2-4$~Gyr). 
The bulge has solar stellar metalicity and a slightly older age of $6-7$~Gyr (I.~Katkov, private communication). 
The \HI rotation curve of \N5533 is declining \citep{Noordermeer2007}, which results in a small radial scale of the dark halo \citep{NoordermeerPhDT}. 
The complex HSB+LSB structure and a short radial scale of the dark halo suggests that the extended LSB disc of \N5533 (similarly to \U1378) could have been formed by the accretion of gas onto a preexisting early-type spiral. 

Another extended LSB system is \N5383 with the disc scale length of 9.7~kpc \citep{vanderKruit1978}. 
This galaxy has a strong bar and extended LSB spiral structure similar to that of \N7589. 
The system also has a companion \U8877 at the distance of roughly 30~kpc with the velocity difference of 100~\kms. 
We see the bifurcation of the spiral arms of \N5383. 
Tidal forces could impose a new mode of the density wave in the disc, which gave the rise to the faint spiral arms like, for example, in the interacting system Arp~82 \citep{Zasov2019}.

Systems \revtwo{morphologically} similar to \U1382, but with smaller discs can be found in the deep images from the MATLAS project \citep[Mass Assembly of early-Type GaLAxies with their fine  Structures][]{Duc2015}. 
One example is \U9519, a dwarf S0 galaxy surrounded by an extended LSB spiral disc with a radius of $\sim$30~kpc. 
\citet{Sil'chenko2019} found out that gas in the inner part of \U9519 rotates in a nearly polar plane with respect to the stellar disc and the outer disc is decoupled from the main body of the galaxy. This could indicate that gas was accreted from external sources \citep{Sil'chenko2019}. 

Starforming gLSBGs can be considered as an extension of the class of galaxies with extended ultraviolet discs \citep[XUV discs][]{GildePaz2005, Thilker2005, Thilker2007} to larger disc sizes and thus could have evolved in a similar way.
\citet{Hagen2016} classified \U1382 as a Type~I XUV disc similar to the prototypical XUV disc galaxies M~83 and \N4625  \citep{GildePaz2005, Thilker2005}. 
\citet{Boissier2016} also notes that \Malin1 could be an extreme case of an antitruncated XUV disc. 
According to \citet{Thilker2007}, Type~I XUV discs are found in upto 20 per cent of galaxies and span the entire range of Hubble types. 
The global characteristics of galaxies with XUV discs are similar to those of `normal' galaxies, which can indicate that the formation of an XUV disc could be a stage of `normal' galaxy formation process. 
\citet{Thilker2007}  pointed out that an interaction could trigger the formation of an XUV disc, or, in other words, trigger a burst of star formation in the extended gaseous disc which would have remained passive and `dark' otherwise.
Another possible trigger may be a high specific rate of gas accretion (i.e. per unit stellar mass).

It is interesting to compare gLSBGs with the famous Hoag's Object, an unusual giant ring galaxy in a low-density environment discovered by \citet{Hoag1950}. 
The blue and young ring surrounds an elliptical galaxy hosting old metal-rich stellar population \citep{Finkelman2011}. 
The radius of the ring is about 25~kpc and it has a clearly visible spiral-like pattern. 
\citet{Finkelman2011} proposed that the peculiar structure and observed properties of the Hoag's Object are the result of the gas accretion onto a preexisting elliptical galaxy. 
The Hoag's Object could thus be a special case of a ``failed'' gLSBGs in which instead of a giant LSB disc, the star-forming ring was formed due to some specific properties of the system, e.g. a triaxial potential of the elliptical galaxy, which could lead to the prominent gap between the core and the ring. 

There exist LSB-systems with complex HSB+LSB structures like that in many gLSBGs but with more moderate sizes. Interesting example of such system is Ark~18 residing in the Eridanus void with an extended blue LSB disc and a bright central elliptically-shaped part  which is probably the result of a dwarf-dwarf merger \citep{Egorova2021}.

\subsection{Comparison of gLSBGs with oversized HSB galaxies}

\citet{Ogle2016} found out that about 6 per cent of the most optically luminous galaxies have giant HSB discs with isophotal diameters of $55-140$~kpc. 
This sample was extended by \citet{Ogle2019} which also included non-star-forming `super-lenticulars' in addition to superspirals and giant elliptical galaxies. 
A small fraction of the superluminous galaxies had disturbed morphology indicating recent mergers. 
Similarly to gLSBGs, super disc galaxies are red inside and blue outside which could be consistent with the on-going growth of the disc by accretion and minor mergers \citep{Ogle2019}. \citet{2020MNRAS.493.5464K} studied in detail the edge-on `super-lenticular' \N7572 with an excessively massive thick disc and concluded that the thin disc growth was likely stopped prematurely by the dense cluster environment. Otherwise, this object would have become an enormous `superspiral' if it could continue to grow its disc from an external gas supply.

\citet{Saburova2018} also studied giant HSB discy galaxies at lower redshifts compared to those from the sample of \citet{Ogle2016} and using different selection criterion, the isophotal radius rather than luminosity used in \citet{Ogle2016}. 
\citet{Saburova2018} derived the parameters of dark haloes of giant HSB galaxies and concluded that they tend to have high halo masses and radial scales in agreement with their large disc radii. 
Thus, \citet{Saburova2018}  proposed that the size of HSB giants is due to the sparse dark halo being similar to the formation scenario by \citet{Kasparova2014} proposed for the gLSBG \Malin2. 

An interesting case is the giant spiral galaxy LEDA\,1970716 discovered by visual inspection of DECaLS optical images. 
It has an asymmetric spiral arm with blue clumps of star formation which extends out to 120~kpc from the centre, whose formation could have been induced by the interaction with neighbouring galaxies. 
This example demonstrates the possibility of the giant LSB disc formation by gravitational interactions (flybys).

\section{Summary}\label{conclusion}

We collected long-slit spectral observations for four gLSBGs galaxies in addition to the three systems with the data available in the literature. We also performed surface photometric analysis for \N7589 and \U6614 on deep archival images. We analysed all available information for the sample of seven gLSBGs and compared them to  galaxies of moderate sizes and to giant HSB galaxies (`superspirals').

Observational data favor the external origin of the giant LSB discs for most gLSBGs. For \N7589, \U1378, and \U1382, we argue for the two-stage formation scenario in which the extended LSB disc is a result of gas accretion on a preexisting early-type galaxy. There also exist gLSBGs like \U1922 which were likely formed by a major merger. For \Malin1 and \U1382, we also can not exclude the major merger hypothesis. Some alternative formation scenarios are also feasible. For \Malin2,  \U6614 and possibly \N7589, we found high radial scale and low central density of their dark haloes, so the unusual properties of their discs could be dictated by the sparse dark matter haloes. 

The proposed formation scenarios are supported by the following observed properties of gLSBGs:
\begin{itemize}
    \item We detected a counterrotation of ionized gas in the inner parts of two out of seven considered gLSBGs. 
    Similar high statistical frequency of the kinematically decoupled kinematics is observed in isolated lenticular galaxies \citep{Katkov2014}, where the external origin of gas via accretion is a preferred scenario.
    
    \item The SFR surface densities of gLSBGs are too low for their gas content, which can be a result of the inefficient star formation, e.g. because of the low gas volume density or very recent accretion of gas. 
      
    \item At least six out of seven presented gLSBGs host active galactic nuclei. Such high AGN rate requires the presence of gas supply reaching the central regions of the galaxies. At the same time, the central black hole masses appear to be significantly lower than expected for the obtained central stellar velocity dispersions, in agreement with previous findings by \citet{Ramya2011, Subramanian2016}. This can indicate that  there were not many major merger events in the history of gLSBGs similar to galaxies with smaller bulges hosting active intermediate-mass black holes \citep{2018ApJ...863....1C}.
      
    \item We performed the mass modelling of optical + \HI rotation curves of the seven gLSBGs using the Burkert dark matter density profile. 
    The derived parameters of dark haloes show different behavior compared to the LSB disc radius and $R_{opt}$ which can indicate different formation scenarios of the considered galaxies.
    
    \item Stellar populations in the central parts of most of the presented gLSBGs are old and metal-rich arguing for a two-stage gLSBG formation.

\end{itemize}

We conclude that the seven presented gLGBGs do not form a homogeneous class of objects and one should consider several alternatives for their formation scenarios, most of which require an external origin of material to form gLSB discs.

\section*{Data availability}
 
The data underlying this paper will be shared upon request to the corresponding author.

\section*{Acknowledgements} We thank the anonymous referee for the valuable and encouraging comments.
We are grateful to Anatoly Zasov, Fran\c coise Combes, and Fr\'ed\'eric Bournaud for fruitful discussion. 
The spectral data reduction and interpretation of the results were supported by the Russian Science Foundation (RScF) grant No. 19-12-00281. The mass modelling of the rotation curves was done with the support of  the Russian Science Foundation (RScF) grant No. 19-72-20089. IC's research is supported by the Telescope Data Center of the Smithsonian Astrophysical Observatory. We acknowledge the usage of the HyperLeda database (\url{http://leda.univ-lyon1.fr}). This research has been supported by the Interdisciplinary Scientific and Educational School of Moscow University ''Fundamental and Applied Space Research''. KG acknowledges the support from the Foundation of development of theoretical physics and mathematics `Basis' (category: students). Observations conducted with the 6-m telescope of the Special Astrophysical Observatory of the Russian Academy of Sciences carried out with the financial support of the Ministry of Science and Higher Education of the Russian Federation (including agreement No. 05.619.21.0016, project ID RFMEFI61919X0016). Based on observations obtained with MegaPrime/MegaCam, a joint project of CFHT and CEA/DAPNIA, at the Canada-France-Hawaii Telescope (CFHT) which is operated by the National Research Council (NRC) of Canada, the Institut National des Science de l'Univers of the Centre National de la Recherche Scientifique (CNRS) of France, and the University of Hawaii. This study is based in part on data collected at Subaru Telescope and retrieved from the Hyper-SuprimeCam data archive system, which is operated by Subaru Telescope and Astronomy Data Center at National Astronomical Observatory of Japan and on observations obtained at the international Gemini Observatory (proposals GN-2005B-Q-61 and GN-2006B-Q-41, data retrieved from the Gemini Science Archive, \url{http://archive.gemini.edu/}), a program of NSF’s NOIRLab, which is managed by the Association of Universities for Research in Astronomy (AURA) under a cooperative agreement with the National Science Foundation on behalf of the Gemini Observatory partnership: the National Science Foundation (United States), National Research Council (Canada), Agencia Nacional de Investigaci\'on y Desarrollo (Chile), Ministerio de Ciencia, Tecnolog\'ia e Innovaci\'on (Argentina), Minist\'erio da Ci\^encia, Tecnologia, Inova\c{c}\~oes e Comunica\c{c}\~oes (Brazil), and Korea Astronomy and Space Science Institute (Republic of Korea). 
The Legacy Surveys consist of three individual and complementary projects: the Dark Energy Camera Legacy Survey (DECaLS; NOAO Proposal ID  2014B-0404; PIs: David Schlegel and Arjun Dey), the Beijing-Arizona Sky Survey (BASS; NOAO Proposal ID 2015A-0801; PIs: Zhou Xu and Xiaohui Fan), and the Mayall z-band Legacy Survey (MzLS; NOAO Proposal ID  2016A-0453; PI: Arjun Dey).
\bibliographystyle{mnras}
\bibliography{saburova}

\label{lastpage}

\end{document}